\documentclass[useAMS,usegraphicx]{mn2e}
\usepackage{amsmath,amssymb}
\usepackage{color,graphicx,colortbl,url}
\usepackage[utf8]{inputenc}
\usepackage[longnamesfirst]{natbib}
\bibpunct[, ]{(}{)}{;}{a}{}{,}

\shortcites{barger_submillimetre-wavelength_1998,barger_redshift_1999,bauer_specific_2005,beswick_evolution_2008,blain_submillimeter_2002,boyle_extendinginfrared_2007,caputi_role_2006,chapman_median_2003,chapman_redshift_2005,chapman_nature_2001,chapman_submillimetre_2002,cohen_deep_2004,cowie_new_1996,daddi_new_2004,dale_infrared_2001,dale_infrared_2005,damen_evolution_2009,de_breuck_sample_2000,de_jong_iras_1984,de_lucia_formation_2006,dickinson_evolution_2003,donley_spitzer_2007,donley_spitzers_2008,dye_scuba/spitzer_2007,eales_canada-uk_1999,eales_canada-uk_2000,feulner_connection_2005,garn_relationship_2009,giovannoli_population_2010,glazebrook_imaging_1995,gott_median_2001,greve_1200-mum_2008,groves_modelingpan-spectral_2008,huynh_infrared_2007,ibar_deep_2009_ii,ivison_diversity_2000,ivison_hyperluminous_1998,kovcs_sharc-2_2006,lilly_canada-france_1996,lutz_nature_1998,madau_high-redshift_1996,magnelli_0.4_2009,maraston_evidence_2006,miller_vla_2008,muzzin_well-sampled_2010,oliver_specific_2010,rieke_determining_2009,roche_atlas_1991,rovilos_radio_2007,sajina_1-1000m_2006,scoville_high-mass_2001,seymour_investigatingfar-ir/radio_2009,seymour_comoving_2010,silva_modelingeffects_1998,smail_nature_2002,smail_radio_2000,symeonidis_link_2009,thompson_magnetic_2006,vega_modellingspectral_2008,virani_extended_2006,wall_parkes_2005,white_signalsnoise:_2007,williams_detection_2009,wuyts_fireworks_2008,zheng_detecting_2006,zheng_infrared_2007}

\arrayrulecolor{black}

\newcommand{\degree}{\ensuremath{^\circ}}
\newcommand{\mum}{\ensuremath{\mu\text{m}}}
\renewcommand{\thefootnote}{\fnsymbol{footnote}}

\title[Evolution of the FIR--Radio Correlation and IR SEDs]{Evolution of the Far-Infrared--Radio Correlation and Infrared SEDs of Massive Galaxies over \textit{z} = 0 -- 2}
\author[N.\ Bourne et al.]
       {N.\ Bourne,$\!^1$\footnotemark\ 
	L.\ Dunne,$\!^1$  
        R.\,J.\ Ivison,$\!^{2,3}$ 
	S.\,J.\ Maddox,$\!^1$ 
	M.\ Dickinson$^4$ and 
	D.\,T.\ Frayer $\!^5$
	\vspace*{1mm}\\
        $^1$ School of Physics \& Astronomy, University of Nottingham,
             University Park, Nottingham NG7 2RD\\
        $^2$ UK Astronomy Technology Centre, Royal Observatory, Blackford
             Hill, Edinburgh EH9 3HJ\\
        $^3$ Institute for Astronomy, University of Edinburgh, Blackford Hill,
             Edinburgh EH9 3HJ\\
        $^4$ National Optical Astronomy Observatory, Tucson, AZ 85719, USA\\
        $^5$ National Radio Astronomy Observatory, PO Box 2, Green Bank, WV 24944, USA
	}
\begin{document}
\maketitle

\begin{abstract}
We investigate the far-infrared--radio correlation (FRC) of
stellar-mass-selected galaxies in the Extended Chandra Deep Field South 
using far-infrared imaging from \textit{Spitzer} and radio imaging from the
{Very Large Array} and {Giant Metre-Wave Radio Telescope}.
We stack in redshift bins to probe galaxies below the noise and confusion limits.
Radio fluxes are $K$--corrected
using observed flux ratios, leading to tentative evidence for
an evolution in spectral index.
We compare spectral energy distribution (SED) templates of local galaxies
for $K$--correcting FIR fluxes, and show that the
data are best fit by a quiescent spiral template (M51) rather than a
warm starburst (M82) or ULIRG (Arp220), implying a predominance of
cold dust in massive galaxies at high redshift.  In contrast we measure
total infrared luminosities that are consistent with high star-formation rates.
We observe that the FRC index ($q$) does not evolve
significantly over $z=0-2$ when computed from $K$-corrected 24 or
160-\mum\ photometry, but that using 70-\mum\ fluxes leads to an apparent
decline in $q$ beyond $z\sim1$.  This suggests some change in
the SED at high redshift, either a steepening of the spectrum at rest-frame
$\sim25-35$\mum\ or a deficiency at $\sim70$\mum\ leading to a drop in the 
total infrared/radio ratios.  We compare our results to other work in the 
literature and find synergies with recent findings {on the high-redshift
FRC, high specific star-formation rates of massive galaxies and the cold dust 
temperatures in these galaxies}.

\end{abstract}
\footnotetext{E-mail: ppxnb1@nottingham.ac.uk}
\renewcommand{\thefootnote}{\arabic{1}}

\section{Introduction}

One of the most exciting research areas in observational astronomy at
this time is the field of far-infrared astronomy.  Our understanding
of extragalactic sources in the far-infrared (FIR) and sub-millimetre
(sub-mm) to millimetre regimes has improved exceptionally over the past
decade, thanks to such instruments as ESA's \textit{Infrared Space
Observatory} (\textit{ISO}, launched 1995;
\citealp{kessler_infrared_1996}) and NASA's \textit{Spitzer Space
Telescope} (launched 2003; \citealp{werner_spitzer_2004}), alongside
ground-based instruments such as the Sub-millimetre Common-User
Bolometer Array (SCUBA; \citealp{holland_scuba:common-user_1999}) and
the Max Planck Millimeter Bolometer Array (MAMBO;
\citealp{kreysa_bolometer_1998}), both commissioned in the late
1990's.  One of the most important FIR instruments to date is the
Multiband Imaging Photometer for \textit{Spitzer}
\citep[MIPS;][]{rieke_multiband_2004}, which, with \textit{Spitzer}'s
0.85m mirror provides imaging with diffraction limited resolutions of
6, 18 and 40 arcsec in three broad bands centred at 24, 70 and
160\mum\ respectively.

Recent advances in the resolution of the cosmic infrared background
(CIB) by stacking into images from \textit{Spitzer} {\citep[see
e.g.][]{dole_cosmic_2006,dye_scuba/spitzer_2007,chary_new_2010}} and BLAST
\citep[The Balloon--borne Large Aperture Sub-mm
Telescope;][]{marsden_blast:_2009} have enabled an improved
understanding of the history of star formation and galaxy formation
and evolution, and this will be further improved by ongoing work 
\citep[such as][]{berta_dissecting_2010} with ESA's new
\textit{Herschel Space Observatory} \citep{pilbratt_herschel_2010}.

Most of the stellar mass in the local universe is
concentrated in the most massive galaxies ($M_\star \gtrsim
10^{11}\text{M}_\odot$; \citealp{kauffmann_stellar_2003}) and observations
show that these have been in place since $z\sim 1$
\citep{dickinson_evolution_2003, bundy_mass_2005,
prez-gonzlez_stellar_2008, taylor_rise_2009, collins_early_2009}.  The
significant increase in density of luminous ($L_{8-1000\mum} \gtrsim
10^{11} \text{L}_\odot$) and ultra-luminous ($L_{8-1000\mum} \gtrsim 10^{12}
\text{L}_\odot$) infrared galaxies (LIRGS and ULIRGs) from the local universe
to $z\sim 2-3$ \citep[e.g.][]{daddi_passively_2005,
daddi_population_2005, caputi_role_2006} is thought to reveal the
formation stages of these latter-day giants, which apparently formed
in a remarkably short time between $1 \lesssim z \lesssim 3$, in an
antithesis to the paradigm of hierarchical structure formation
\citep[e.g.][]{de_lucia_formation_2006}.

\label{sec:sources}
In the FIR--sub-mm regime the dominant source of the extragalactic
background light (after the cosmic microwave background) is thermal
continuum emission from interstellar dust, which is mainly composed of
polycyclic aromatic hydrocarbons (PAHs), graphites and silicates,
typically less than $\sim 0.25\mum$\ in size
\citep{draine_infrared_2007, draine_dust_2007}.  The FIR emission in
star-forming galaxies is thought to arise both from cold dust in the
large-scale `cirrus' component of the interstellar medium (ISM), and
from warmer dust in and around star-forming regions
\citep[e.g.][]{de_jong_iras_1984,helou_iras_1986}.  If there is a
sufficient level of dust-enshrouded star formation then the FIR
emission is dominated by this `warm' dust component, which has
characteristic temperatures of around 30--50K \citep{dunne_scuba_2001,
sajina_1-1000m_2006, dye_scuba/spitzer_2007,
pascale_blast:far-infrared_2009}.  The dust is heated to these
temperatures by the ultraviolet (UV) radiation field from hot O and B
type stars, which are present only in regions of ongoing or recent
star formation (`recent' meaning within the lifetime of these
short-lived stars, $\lesssim10^{8}\text{yr}$; \citealp{kennicutt_star_1998}).
For this reason, the total IR luminosity ($L_\text{TIR} = L_{8-1000\mum}$) can be used
as a tracer of star-formation rates (SFRs) in galaxies, often using
one or a number of FIR fluxes (such as the \textit{Spitzer} MIPS
bands) or sub-mm fluxes to estimate $L_\text{TIR}$
\citep{kennicutt_star_1998}.

One issue with using the FIR as a SFR tracer is the contribution from
cold dust in the ISM, which is heated by older stars in the disk of
the galaxy, and is therefore unrelated to star formation \citep{calzetti_calibration_2010}.
Contamination of samples by galaxies hosting active galactic nuclei
(AGN) is another problem, as AGN also emit UV radiation which can heat
dust in the torus.  AGN-heated dust is generally hotter than dust
heated in star-forming regions, so the thermal spectrum peaks at a
shorter wavelength, and mid-infrared (MIR) fluxes (including
\textit{Spitzer}'s 24-\mum\ band, as well as shorter wavelengths) would
be boosted.  Mid-infrared fluxes are also affected more uncertainly by
the emission features of PAH molecules, which are ubiquitous in
star-forming galaxies \citep[e.g.][]{leger_identification_1984,
roche_atlas_1991, lutz_nature_1998, allamandola_modeling_1999}, and
the 10-\mum\ silicate absorption trough, so the use of 24-\mum\ fluxes
as SFR indicators at high redshifts is subject to some contention
\citep[see e.g.][]{dale_infrared_2005, calzetti_calibration_2007,
daddi_multiwavelength_2007, papovich_spitzer_2007, young_mid-_2009,
rieke_determining_2009}.

\label{sec:FIRRC}
Another part of a galaxy's SED that can be used as a SFR indicator is
the radio luminosity.  Non-thermal radio continuum emission
from star-forming galaxies originates from type II supernova remnants
(SNRs), the endpoints of the same massive short-lived stars that heat
the dust via their UV radiation.  This connection with dust heating is
important because it leads to the well-known (but not fully
understood) FIR--Radio correlation
\citep[FRC;][etc]{van_der_kruit_high-resolution_1973,
rickard_far-infrared_1984, helou_thermal_1985, condon_radio_1992}.
The FRC is linear, remarkably tight and holds for a wide range of
galaxy types over at least five orders of magnitude in luminosity
\citep{yun_radio_2001}.  It can be explained in terms of ongoing
star~formation producing hot massive ($M>8\text{M}_\odot$) stars: while the
FIR flux is emitted from dust heated by these stars, the radio
emission arises from synchrotron radiation by cosmic ray (CR)
electrons accelerated in the SNRs of the dying stars.  The non-thermal
radio emission is smeared out through the galaxy as the relativistic
CR electrons travel through the galaxy over lifetimes of $\tau \sim
10^{8}$ years, during which they emit synchrotron radiation via
interactions with the galactic magnetic field
{(as described by \citealt{condon_radio_1992}, and shown observationally by 
\citealt{murphy_effect_2006})}.  A shallower thermal component is also
present in the radio spectrum due to bremsstrahlung radiation from
electrons in H\,{\sc ii} regions, but this becomes dominant only at
high frequencies ($\gtrsim 30$~GHz) and at 1.4~GHz only comprises 
$\sim 10\%$ of the radio flux \citep{condon_radio_1992}.

It is difficult however to explain the linearity and tightness of the
correlation between thermal FIR luminosity and non-thermal radio
luminosity.  `Minimum energy' estimates of magnetic fields in galaxies
\citep{burbidge_synchrotron_1956}\footnote{
Inferring magnetic field strengths of galaxies is problematic and depends on many unknowns.
The minimum energy argument circumvents some of these problems by specifying the minimum
of the total energy density (a function of field strength), which occurs at the point where
the energy density in particles is in approximate equipartition with that in the magnetic field
\citep{longair_high_1994}.
Although there is no physical requirement for this equipartition to occur, it does provide an order-of-magnitude
estimate for the magnetic field energy density, and is physically motivated if one considers that the magnetic 
field is tangled by turbulent motions in the interstellar plasma, and that these motions efficiently accelerate
cosmic ray particles (see \citealt{longair_high_1994} and \citealt{thompson_magnetic_2006}).
Equipartition has been observationally confirmed locally in the Milky Way \citep{strong_diffuse_2000}
and in normal spiral galaxies \citep{vallee_galactic_1995,beck_magnetic_2000},
although its validity in starbursts and ULIRGs is less certain \citep{thompson_magnetic_2006}.} 
imply a large variation between normal galaxies and extreme starbursts
like Arp220.  To explain a constant FIR/radio ratio between such
disparate systems, complex physical solutions need to be provided, for
example invoking strong fine-tuning to regulate electron escape and
cooling timescales, or short cooling timescales with magnetic fields
$\sim10$ times stronger than the `minimum energy' argument suggests
\citep{thompson_magnetic_2006}.

\Citet{voelk_correlation_1989} 
first suggested a `calorimeter' model whereby both UV light from
massive stars and CR electrons from SNRs are (a) proportional to the
supernova rate, and (b) efficiently absorbed and reprocessed within
the galaxy, so that the respective energy outputs in FIR re-radiation
and radio synchrotron would both be tied to the supernova rate.  This
theory requires a correlation between the average energy density of
the radiation field and the galaxy magnetic field energy density.
\Citet{voelk_correlation_1989} argues this is plausible if the origin
of the magnetic field is a turbulent dynamo effect, since the
turbulence would be largely caused by the activity of massive stars,
and hence correlated with the supernova rate.  Alternative
non-calorimetric models include those of \citet{helou_physical_1993},
using a correlation between disk scale height and the escape scale
length for CR electrons; and \citet{niklas_new_1997}, in which the FRC
is driven by correlations with the overall gas density and
equipartition of magnetic field and CR energy.

\Citet{bell_estimating_2003} argues for a `conspiracy' to diminish both the FIR and radio emission
originating from star formation in low luminosity galaxies when
compared with luminous $\sim L_\star$ galaxies -- without this the
relationship would not remain linear over the full luminosity range.
Similarly, the calorimeter model of \citet{lacki_physics_2009-1}
invokes conspiracies in low and high gas surface density regions to
maintain the relationship.  The physical origin of the FRC therefore
is still an open question.  A full review of the theories is beyond
the scope of this study, but a more detailed discussion of the
literature can be found in \citet{vlahakis_far-infrared-radio_2007},
and a more in-depth treatment is provided {in the numerical work of}
\citet{lacki_physics_2009-1}.

An ongoing strand of research at the current time is the investigation
of the FRC at high redshifts and low fluxes, in
particular whether there is any evolution \citep[e.g.][]{
appleton_far-_2004, frayer_spitzer_2006,
ibar_exploringinfrared/radio_2008, garn_relationship_2009,
seymour_investigatingfar-ir/radio_2009,
ivison_blast:far-infrared/radio_2009,sargent_vla-cosmos_2010,sargent_no_evolution_2010}.  
Measurements of any evolution (or lack thereof) would improve the accuracy of FIR-/radio-estimated
SFRs at high redshift, and could shed light on the mechanism governing
the FRC, as well as highlighting differences in the physical and
chemical properties of star-forming galaxies at high and low redshift
\citep{seymour_investigatingfar-ir/radio_2009}.

In the current work we investigate the FRC over a range of redshifts,
for a sample that is not limited by FIR or radio flux.  Using
\textit{Spitzer} FIR data and radio data from the {Very Large
Array} ({VLA}) and {Giant Metre-Wave Radio Telescope}
({GMRT}), we quantify the FIR--Radio Correlation as a function
of redshift in massive galaxies selected from a near-infrared (NIR)
survey of the Extended Chandra Deep Field South (ECDFS).  We use
Equation~\ref{eqn:q_ir} to define the `$q$' index, which quantifies
the FRC as the logarithmic ratio between a monochromatic FIR flux 
($S_{\nu,\text{IR}}$, e.g. at 24, 70 or 160\mum), and 1.4-GHz radio flux
($S_{\nu,\text{1.4~GHz}}$).
\begin{equation}
 \label{eqn:q_ir}
 q_\text{IR} = \log_{10}\left(\frac{S_{\nu,\text{IR}}}{S_{\nu,\text{1.4~GHz}}}\right)
\end{equation}
We also investigate the effects of using different FIR bands to
quantify the FRC, and the effects of assumptions about the SEDs of the
galaxies in the sample.  We employ a `stacking' methodology to recover
sufficient signal-to-noise ratios on faint objects to obtain
measurements of the average properties of the sample.  The data are
described in Section~\ref{sec:data}, while the binning and stacking
methodologies are described in Section~\ref{sec:method}.  The analysis
of SEDs and application of $K$--corrections is covered in
Section~\ref{sec:kcorrs}, and the results are analysed and discussed
in Section~\ref{sec:discussion}.  A concordance cosmology of
$\Omega_m=0.27$, $\Omega_\Lambda=0.73$, $H_0=71$\ kms$^{-1}$Mpc$^{-1}$
is assumed throughout.

\section{Data}
\label{sec:data}

The ECDFS is a $\sim\,0.25$\ deg$^2$ square centred at
$3^h32^m30^s$,~$ -27 \degree 48'20''$ (J2000).  It is a much-studied region of
sky, with a rich body of published data and studies of extragalactic
sources at a broad range of wavelengths stretching from X-ray to radio
regimes.  We use radio imaging data at 1.4~GHz from the
{VLA}, with a typical rms sensitivity across the map of $8\mu \text{Jy beam}^{-1}$ 
and beam dimensions of $2.8 \times 1.8$ arcsec \citep{miller_vla_2008}. 
Imaging data were obtained at 610~MHz from the {GMRT}, reaching a sensitivity of 
$\sim 40 \mu \text{Jy beam}^{-1}$ with a $6.5 \times 5.4$ arcsec beam 
\citep[][]{ivison_blast:far-infrared/radio_2009}. 
For the FIR, \textit{Spitzer} MIPS images at 24, 70 and 160\mum\ were obtained from
the FIDEL survey (DR3; Dickinson et al., in preparation).

To look at a range of galaxy types over a range of redshifts we must
give careful thought to how the galaxies are selected.  For
example, selecting radio-bright galaxies will naturally favour active
radio galaxies, while selection at 24\mum\ is likely to favour
galaxies with dusty starbursts and/or obscured AGN components.  These
biases will affect the measured value of $q_\text{IR}$ \citep[see
e.g.][]{sargent_vla-cosmos_2010}.  There is however a good body of
evidence that distant massive galaxies in a range of phases of
star-formation and nuclear activity can be effectively selected in NIR
filters at $\sim 2$\mum.  This part of the spectrum is minimally
affected by dust absorption, AGN and other components, {and} hence is
relatively insensitive to the `type' of galaxy or the shape of its
SED.  Furthermore it is insensitive to the age of the stellar
population (hence SFR), because the light is dominated by old
main-sequence stars that make up the bulk of the stellar mass in all
galaxies. Thus NIR luminosity is primarily dependent on stellar mass
only \citep{glazebrook_imaging_1995, gardner_near-infrared_1995}.

In the ECDFS there exists NIR data from the \textit{Spitzer} Infrared
Array Camera \citep[IRAC;][]{fazio_infrared_2004}, of which the two
shortest wavelength bands (3.6 and 4.5\mum) can also be used as an
effective tracer of stellar mass \citep[e.g.][]{serjeant_scuba_2008}.
A catalogue of IRAC sources matched with optical--NIR 
photometry in the Multiwavelength Survey by Yale--Chile 
\citep[MUSYC;][]{gawiser_multiwavelength_2006} is collected in 
\textit{Spitzer}'s IRAC and MUSYC Public Legacy of the ECDFS 
(SIMPLE; Damen et al., in preparation). The catalogue was 
extracted from IRAC 3.6 and 4.5-\mum\ images, and sources for which the
mean of the 3.6 and 4.5-\mum\ {AB} magnitudes $([3.6]+[4.5])/2<21.2$ were
selected, giving a catalogue of 3841 sources.  The source extraction
and selection is described by \citet{damen_evolution_2009}.  In SIMPLE
the IRAC sources have been matched to multiwavelength counterparts in
the MUSYC catalogue, which contains
photometry in \textit{UBVRIz'JHK} bands.  Stars have been identified
and excluded from the catalogue using the colour criterion $J-K <
0.04$, and potential AGN were removed by excluding any matches with
Chandra X-ray sources \citep{virani_extended_2006}.  Photometric
redshifts were collated for all objects in the sample from COMBO-17
\citep{wolf_catalogue_2004}, and by using the EAZY code
\citep{brammer_eazy:fast_2008} as described by
\citet{damen_evolution_2009}.  Damen et al. compared the photometric
redshifts to spectroscopic ones where available, and showed that the
median $(z_\text{spec}-z_\text{phot})/(1+z)=0.033$ (0.079 at $z\geq1$).  As
described in Section~\ref{sec:sampling}, we divide the sample into
bins with sizes $\Delta z/(1+z) \sim 0.2-0.4$, so it is safe to
neglect these photometric uncertainties.  The final catalogue used in
this work contains 3529 sources with photometric redshifts up to
$z=2$, in the region of the ECDFS defined by the rectangle $52\degree
51'48''~<~$RA$~<~53\degree 25'14''$, $-28\degree
03'27''~<~$Dec$~<~-27\degree 33'22''$ (J2000).

\section{Stacking Methodology}
\label{sec:method}
\subsection{Sample Selection and Redshift Binning}
\label{sec:sampling}

Since sources are selected by their NIR flux across a range of SED
types, many are likely to be faint or undetectable at the wavelengths
of interest.
\footnote{{As an indication of the need to stack, we roughly 
estimated the minimum SFRs detectable at high redshift, using the noise levels in 
Table~\ref{tab:errors} and calculating SFRs that would be derived from 
$5\sigma$ detections using the methodology described in Sections 4 and 5.
These are, from the MIPS and radio fluxes respectively,
870 \& 130$\text{M}_\odot \text{yr}^{-1}$ at $z=1$, and 
4900 \& 870$\text{M}_\odot \text{yr}^{-1}$ at $z=2$.}}
In order to probe the evolution of fluxes as a function
of redshift, we stack galaxies into seven bins in redshift and measure
median fluxes in each bin.  The great advantage of this technique is
the gain in signal-to-noise ratio, as combining many sources reduces
the random noise while maintaining the average level of the signal.
This gain is at the expense of knowledge of the individual galaxies,
but with careful application of criteria when binning the galaxies,
and with a large enough sample, it can reveal properties of galaxies
below the noise and confusion levels.  The technique has been used to
great effect many times in the literature; for example by
\citet{serjeant_submillimeter_2004, dole_cosmic_2006, ivison_aegis20:_2007,
takagi_scuba_2007, white_signalsnoise:_2007, papovich_spitzer_2007,
dunne_star_2009}.

We do not know the distribution of fluxes in the stacks, but since we
select massive galaxies with unknown SEDs at a range of redshifts, we
may expect to be prone to some outliers.  For example, radio-bright
AGN have unusually high radio fluxes and are outliers on the FRC.
{We cannot be certain that these have been successfully removed
from the sample by cross-matching with the Chandra catalogue, 
as we know that there is limited overlap between X-ray and radio-selected 
AGN samples \citep[although see Section~\ref{sec:radio-disc}]{rovilos_radio_2007,
pierce_effects_2010,griffith_morphologies_2010}.}
We therefore used the median statistic to represent the properties of the
typical galaxies in each stack, because unlike the mean, the median is
resistant to outliers \citep{gott_median_2001, white_signalsnoise:_2007, 
carilli_star_2008, dunne_star_2009}.

To study redshift evolution, we divided the sample according to the
photometric redshifts in the catalogue, and stacked each bin into the
radio and FIR images.  The sample was split into the redshift bins
given in Table~\ref{tab:zbins}.

\begin{figure}
\begin{center}
\includegraphics[width=0.45\textwidth]{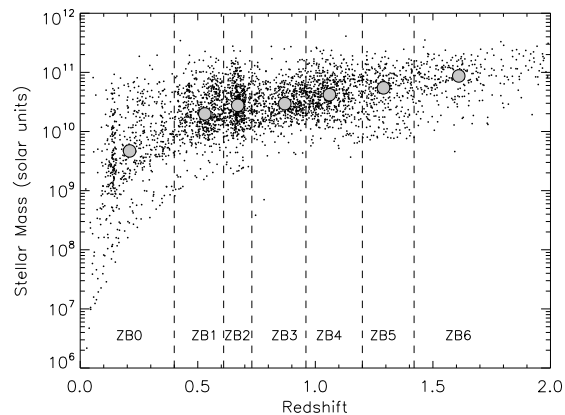}
\end{center}
\caption{Scatter plot of stellar masses in the catalogue as a function of photometric redshift, with the divisions 
between the redshift bins marked as dashed lines.  Large circles mark the median mass and redshift in each bin.}
\label{fig:mass}
\end{figure}

The most significant sampling bias that we expect to see is that of
stellar mass.  The stellar masses of galaxies in the catalogue have
been estimated by \citet{damen_evolution_2009}, by SED-fitting with a
{\citet{kroupa_variation_2001}} initial mass function (IMF).  The effect of Malmquist bias is
that the median mass in the sample is lower at lower redshifts
(because low mass galaxies dominate the population), but is higher at
higher redshifts where only the more massive galaxies are detectable.
This is illustrated in Fig.~\ref{fig:mass}, and we test the effect on
our results in Section~\ref{sec:discussion}.

\begin{table}
\begin{center}
\begin{tabular}[hb]{ c r c l c c }
 \hline
 Bin & \multicolumn{3}{c}{Boundaries} & Median $\langle z \rangle$ & Count $N$ \\ \hline
 ALL   &   0.00 & $\leq z <$ & 2.00 &     0.73     & 3172 \\
 ZB0   &   0.00 & $\leq z <$ & 0.40 &     0.21     & 528 \\
 ZB1   &   0.40 & $\leq z <$ & 0.61 &     0.53     & 529 \\
 ZB2   &   0.61 & $\leq z <$ & 0.73 &     0.67     & 529 \\
 ZB3   &   0.73 & $\leq z <$ & 0.96 &     0.87     & 528 \\
 ZB4   &   0.96 & $\leq z <$ & 1.20 &     1.06     & 529 \\
 ZB5   &   1.20 & $\leq z <$ & 1.42 &     1.29     & 264 \\
 ZB6   &   1.42 & $\leq z <$ & 2.00 &     1.61     & 265 \\ \hline
\end{tabular}
\end{center}
\caption{Redshift bins and statistics of catalogue}
\label{tab:zbins}
\end{table}

\begin{figure*}
\begin{center}
\includegraphics[width=0.6\textwidth]{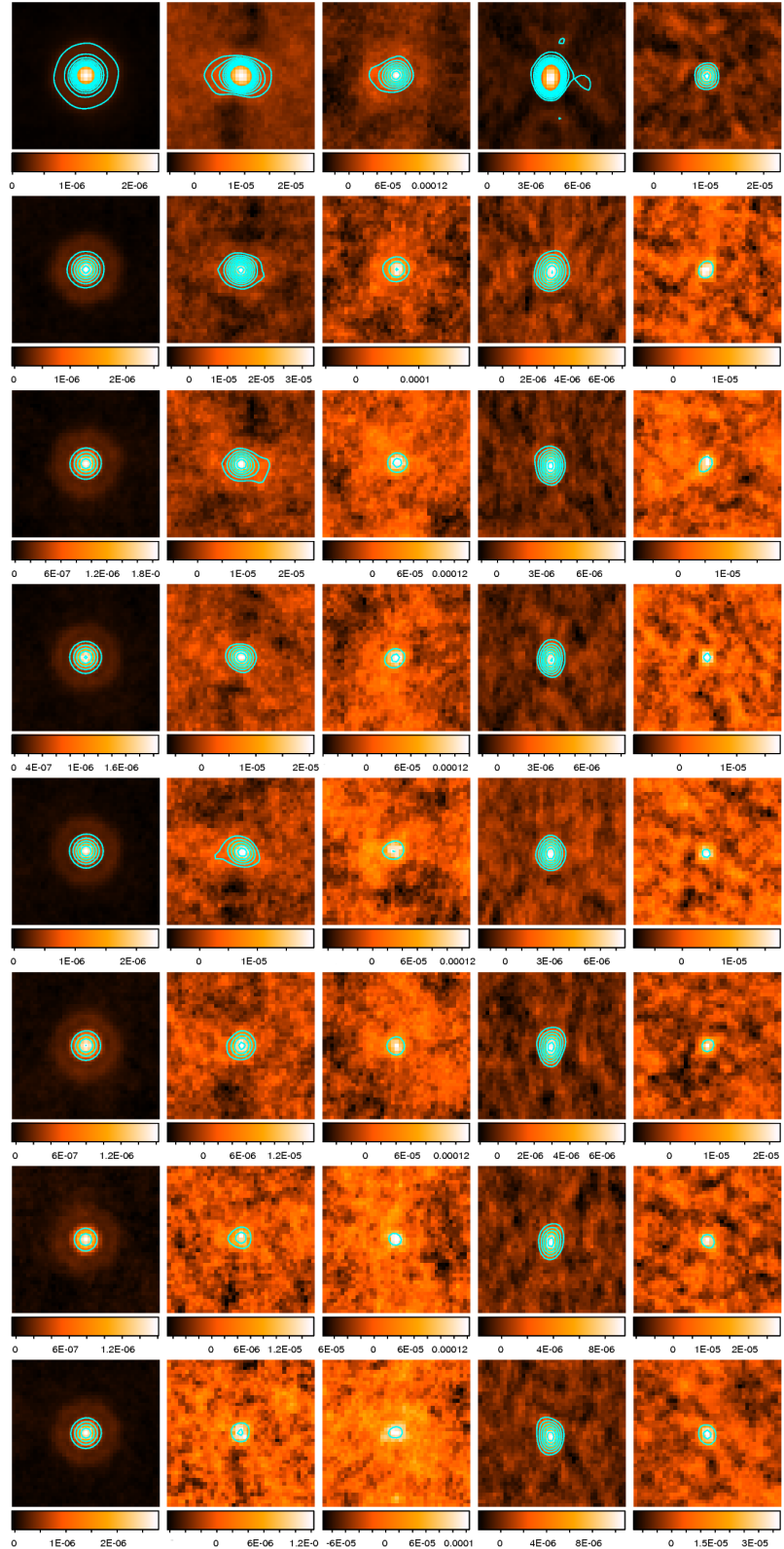}
\caption{Postage-stamp images of the stacked targets in (from left to right) 24\mum, 70\mum, 160\mum, 1.4~GHz, 
610~MHz.  The top row is the stacked full sample; other rows are the stacks of redshift bins ZB0--ZB6 (from top to 
bottom) as described in Section~\ref{sec:sampling}.  Images are coloured by flux in Jy pixel$^{-1}$ units (MIPS) and 
Jy beam$^{-1}$ units (radio), and colour scales are included below each.  Contours of [3, 5, ... 17, 19] times the 
pixel noise are overlaid.  Each tile is a square 41 pixels on a side, which corresponds to 49.2, 164, 328, 20.5 and 
61.5 arcsec for the five bands respectively.}
\label{fig:postagestamps}
\end{center}
\end{figure*}

\subsection{Stacking into the {VLA} and {GMRT} Radio Images}
\label{sec:stackradio}

Radio images have pixel units of Janskys (Jy) per beam, and adhere to
the convention whereby each pixel value is equal to the flux density
of a point source located at that position.  The assumption that the
pixel value at the position of each catalogue object gives the correct
radio flux density for that source is generally good, though requires
a small correction to give the total integrated flux of the source.
This integrated-flux correction accounts for any sources being
extended over more than one beam, and also for any astrometric offset
between the catalogue coordinates and the radio source position.
Since we are stacking, it is suitable to consider the overall effect
on the median stacked source, so the integrated-flux correction is
calculated from the stacked `postage-stamp' image of the full sample.
This image is created by cutting out a 41 pixel square\footnote{41
pixels corresponds to 49.2, 164, 328, 20.5 and 61.5 arcsec in the
24-\mum, 70-\mum, 160-\mum, 1.4-GHz and 610-MHz bands respectively.} 
centred on each source, and stacking the images by taking the median
value of each pixel.  The integrated flux is calculated using the
\textsc{aips} package \textsc{jmfit} and the correction is simply the
ratio of this to {the value of the peak (central) pixel in the image.}

A higher integrated/peak ratio might be expected in the lowest-redshift
bins if a large number of resolved sources were included in
the stack.  We therefore tested whether the correction varied
significantly between different redshift bins, by constraining the
beam centroid position and orientation and measuring integrated-/peak-flux
ratios in each postage stamp.  Without these constraints the
fitted fluxes would be more prone to flux boosting by noise peaks, as
the signal-to-noise in these postage stamps is low.  In both 1.4~GHz
and 610~MHz however, the result for each of the bins was consistent
(within $1\sigma$) with the result for the stack of all sources,
indicating that the stacked sources were unresolved in all bins.
Hence for all bins we used the integrated-flux corrections from the
full stack (which have the smallest errors), i.e. $1.55\pm0.08$ for
1.4~GHz and $1.05\pm0.12$ for 610~MHz.

A further correction would need to be made to point-source fluxes
measured in the 1.4~GHz image to account for bandwidth smearing (BWS),
an instrumental effect caused by the finite bandwidth of the receiver
resulting in sources appearing more extended with increasing angular distance 
from the centre of the pointing.  Since integrated flux is conserved in BWS,
the effect is corrected in our data by the integrated-flux correction
{(this explains the large correction at 1.4~GHz which would otherwise 
appear to be inconsistent with unresolved sources).}

\subsection{Stacking into the \textit{Spitzer} FIDEL Images}
\label{sec:stackfidel}

Measuring fluxes in the MIPS images requires a different technique,
due to the large point-spread function (PSF) which results from the
diffraction-limited resolution of MIPS.  The centre of the PSF can be
described by a roughly Gaussian profile, with full widths at half
maximum (FWHM) of 6, 18 and 40 arcsec in the 24, 70 and 160-\mum\ bands
respectively \citep{rieke_multiband_2004}.  The outer portion of the
24-\mum\ PSF is less predictable and it is known to vary between
different pointings and different source colours\footnote{See MIPS
Data Handbook available at
\url{http://ssc.spitzer.caltech.edu/mips/dh/}}.  For this reason, and
to allow for potentially resolved sources, we chose to measure 24-\mum\
fluxes by aperture photometry, and adopted an aperture of 13 arcsec
with an aperture correction of 1.16 based on the theoretical
\textsc{STinyTim} PRF\footnote{\textsc{TinyTim} for \textit{Spitzer}
developed by John Krist for the \textit{Spitzer} Science Center.  The Center is
managed by the California Institute of Technology under a contract
with NASA.  Web page available at
\url{http://ssc.spitzer.caltech.edu/archanaly/ contributed/stinytim/index.html}}
and the results of \citet{fadda_spitzer_2006},\footnote{Sample xFLS
and \textsc{STinyTim} PRFs are available on the SSC website:
\url{http://ssc.spitzer.caltech.edu/mips/psf.html}} to measure total
fluxes in Jy.  Due to poorer resolution in the 70 and 160-\mum\ maps,
it is sufficiently accurate to measure point-source fluxes by applying
a correction to the central pixel value: the factors used were 43.04
and 46.86 at 70 and 160\mum\ respectively.  This converts
the fluxes to units of Jy beam$^{-1}$, and accounts for large-scale 
emission in the wings of the PSF, as well as a colour
correction.  No further correction is required to measure total
(integrated) fluxes as we can confidently assume that none of the
sources is larger than the beam in these two bands.

In stacking the FIDEL images it was necessary to exclude objects close
to the edges of the map where the noise was higher, {to ensure that
noise in the stacks reduced as $1/\sqrt{N}$} and to prevent
gradients being introduced into the postage stamps.  This was achieved
by placing lower limits on integration time.  Limits were chosen based
on stacks of random positions, resulting in the exclusion of 3.5, 8.6
and 9.9 percent {of the 3529} catalogue sources in the 24, 70 and 160\mum\ bands
respectively.  {Because these cuts are based on integration time alone,
there is no correlation with the nature of the sources themselves, so 
no systematic effect on the measured properties of the galaxies will
be introduced.}

Postage-stamp images of the stacked bins in the MIPS and radio maps
are shown in Fig.~\ref{fig:postagestamps}, including noise contours as
described in the following section.

\subsection{Analysis of Random Errors in Stacked Flux Measurements}
\label{sec:errors}
Random errors in any flux {measurement} arise from noise in the image.
Simplistically these errors might be expected to arise purely from the
variance of pixel values in the map, $\sigma^2$, and the error on $N$
stacked measurements is then given by $\sigma/\sqrt{N}$.  This
assumption is valid for the radio images, so it is sufficient to use
the rms values at the corresponding positions on the rms map.  In the
MIPS data maps however, pixel covariance provides a non-negligible
contribution to the error, so the rms maps are not sufficient.  In
order to measure the total random error on a measured flux we chose
random positions in the sample region of the map and selected those
that fell on empty regions of sky.  This was tested by taking an
aperture of radius 13 arcsec (the radius for aperture photometry at
24\mum) around each position and measuring the standard deviation of
pixel values in that aperture.  If the aperture contained any pixels
that deviated from the aperture mean by more than $2.5\sigma$, then
the position was discarded.  The positions were also required to be
separated and not overlapping.  Thus positions were chosen to
represent regions of empty sky with no sources.  The number of
positions used was chosen to be 500, to roughly match the sample size
of bins used for stacking sources.  For the 24\mum\ case, where
aperture photometry was used, the fluxes at the 500 positions were
measured in exactly the same aperture as was used for source
photometry, and the standard deviation of these sky fluxes was taken
to represent the random error on an individual aperture measurement.
Repeating the analysis with different random catalogues of varying
sizes produced consistent results, and the distribution of flux values
in the random catalogues was {verified} to be Gaussian with high
certainty.

\begin{figure}
\begin{center}
\includegraphics[width=0.45\textwidth]{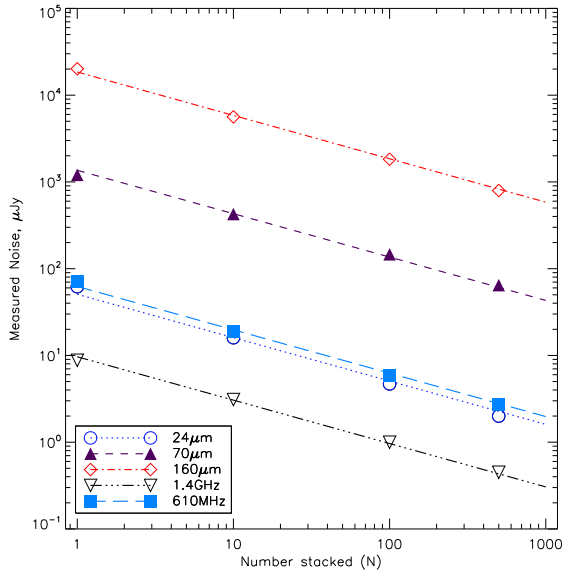}
\caption{{Results of measuring noise in stacks of empty sky positions.
The noise was calculated by stacking $N$ random positions, repeating 
500 times, and taking the standard deviation of the 500 stacked fluxes
(see text for details).
Results of these stacks are shown as symbols, and are in the same units
as in Table~\ref{tab:errors}.  Lines show the least-squares
fit to the results of each band with a power-law index of $-0.5$, representing
the expected reduction of noise as $1/\sqrt{N}$ for each band. 
The agreement between results and expectations is good.
}}
\label{fig:noise}
\end{center}

\end{figure}

The analysis was repeated for the other two MIPS maps and also both
radio maps, using the central {(brightest)} pixel for flux measurement since this
was the method used for source photometry in those bands.  The
resulting error or noise values are given in Table~\ref{tab:errors}.
The radio error values were close to the average value in the rms map,
confirming that the radio error is equal to the rms value.  These
values are then multiplied by the aperture correction or integrated-flux 
correction in each case to represent the error on a flux
measurement in Janskys.  We checked that the noise in a stack reduces
as roughly $1/\sqrt{N}$, {as shown in Fig.~\ref{fig:noise},}
so the errors given in Table~\ref{tab:errors}
are divided by the square root of the number of objects stacked, $N$.

\begin{table*}
\begin{center}
\begin{tabular}{ l c c c c }
\hline 
Band & Pixel scale & PSF/Beam FWHM & Noise level & Background level \\  
     & arcsec & arcsec & $\mu$Jy     & $\mu$Jy \\ \hline
24\mum  & 1.2 & 5.9            &     62          & $ -37.60  \pm 0.04 $ \\
70\mum  & 4.0 & 18             &  1,200          & $ +2.2    \pm 0.8 $   \\
160\mum & 8.0 & 40             & 20,000          & $ +2,000  \pm 10 $   \\
1.4~GHz  & 0.5 & $2.8\times1.5$ &   8.83          & $ -0.014  \pm 0.005 $  \\
610~MHz  & 1.5 & 7.7            &   71.9          & $ -0.01   \pm 0.03$  \\
\hline

\end{tabular}
\end{center}
\caption{Information on the FIDEL and radio images.  Measured total
noise values represent the {$1\sigma$} error on a single flux measurement; for
24\mum\ this is the noise on a corrected aperture flux, for 70 and
160\mum\ it is the noise on a corrected point source flux, and for the
radio it is the noise in a beam and does not include the integrated-flux 
correction.  Background levels are in the same units, these are
the values subtracted from the median source fluxes.  Errors on
background fluxes are standard errors from 1000 measurements as
described in Section~\ref{sec:bgs}.}
\label{tab:errors}
\end{table*}

In using the median to represent the fluxes of $N$ sources
in some bin, we must also consider the width of the distribution of fluxes
in that bin: if this is larger than the estimated measurement error$/\sqrt{N}$
then the latter is a poor indicator of uncertainty on the quoted median.
For this reason we estimated $1\sigma$ uncertainties on the median 
following the method of \citet{gott_median_2001} and compared them to the
estimated measurement error in each stack (the value in Table~\ref{tab:errors}
divided by $\sqrt{N}$).  At 160\mum, 610~MHz and 1.4~GHz we found the two to be 
about equal  (see Table~\ref{tab:stackresults}), confirming that the flux
errors we have estimated cover the distribution of fluxes in the bins.
At 24\mum\ the uncertainty on the median was around three times the size of the 
estimated flux error, and at 70\mum\ around twice the size, indicating that in
these bands the flux distribution in each bin was somewhat broader than the
estimated errors allowed for.  It is possible that the method described 
in preceding paragraphs systematically underestimates the noise in these images,
as a result of the constraints used to identify empty `sky' apertures.  Those
constraints were designed to distinguish true read-noise on the detector from confusion 
noise in the sky, but the 24-\mum\ image in particular is highly confused, meaning
that the constraints coud lead to correlation in the empty apertures stacked, and increase the
chance of underestimating the noise.
For the analysis of stacked results in this paper we therefore quote the 
uncertainties on the median following the \citet{gott_median_2001} method.

\subsection{Background Subtraction and Clustering Analysis}
\label{sec:bgs}

\begin{figure}
\begin{center}
\includegraphics[width=0.45\textwidth]{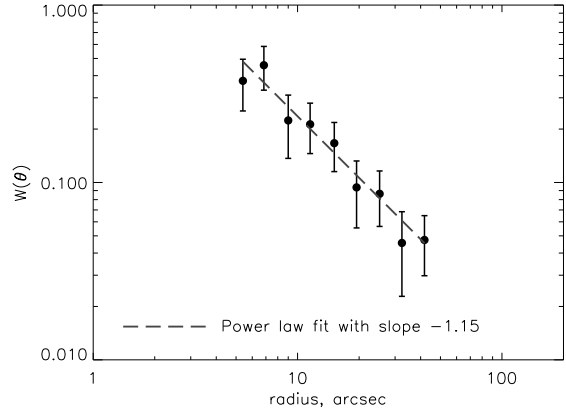}
\caption{Autocorrelation function $W(\theta)$ of the SIMPLE catalogue.  
The dashed line is a power-law fit with index -1.15.  Error bars are twice the Poisson errors, 
as described in the text.}
\label{fig:wtheta}
\end{center}
\end{figure}

A similar methodology to the random error analysis was used to measure the background value to be
subtracted.  The method described above chooses empty apertures
containing just sky, which is simplistically what needs to be
subtracted before performing aperture photometry, but in the case of
MIPS the combination of high source density and low resolution require
that source confusion is also accounted for.  When stacking, the
random boosting of fluxes on individual sources will average out to a
constant correction that can be included in the background.  Hence
when measuring the background for subtraction, a catalogue of random
positions were chosen and stacked, without any criteria on the
existence or otherwise of sources close to these random positions.  On
average the random catalogue should coincide with sources with the
same probability as the source catalogue does.  
{In this context we use the term `sources' in the general sense, meaning
any source of flux in the images that we stack into, i.e. any object that could
boost a measured flux at a given position.}  At this stage we are
making the assumption that source clustering does not play a part.
Stacks of 3500 random positions (to match the sample size of the
source catalogue) were made and repeated 1000 times in each of the
three MIPS and two radio maps.  The mean of the 1000 stacked fluxes
was taken to be the background value, and the standard error was taken
to be the uncertainty; results are given in Table~\ref{tab:errors}.

Any clustering of the sources {in the catalogue} would lead to an increased probability
of confusion for a catalogue source compared with a random position,
hence with increased clustering the background subtraction becomes
increasingly less effective.  In order to estimate the size of this
effect we would ideally need to understand the correlation function of
sources in each image on scales smaller than the beam size.  Since
this is not possible, we made the assumption that the correlation of
sources in the images that we stack into is approximately the same as that in the source
catalogue.  This may not fully account for confusion if the sources in
the image are more clustered than the IRAC {(catalogue)} sources, but it does at
least remove the possibility of double-counting the fluxes of confused
sources in the catalogue.

We calculated the autocorrelation function $W(\theta)$ of positions in
the catalogue, to estimate the excess probability of a background
source appearing at a radius $\theta$ from a target source,
compared with a random position.  We used the \citet{landy_bias_1993}
estimator (Equation~\ref{eqn:landy_szalay}) which counts pairs within
and between the data ($D$) and random ($R$) positions as a function of
annular radius $\theta$.  

\begin{equation}
 W_{D,D}(\theta)=\dfrac{DD - 2\,DR - RR}{RR}
\label{eqn:landy_szalay}
\end{equation}

Results are shown in Fig.~\ref{fig:wtheta}, which includes a fit by
linear regression given by $W(\theta)=0.000269\,\theta^{-1.15}$, where
$\theta$ is in degrees.  {By dividing the region into four eqaul quadrants
and comparing the scatter between results in each, we found that the standard
error was a factor 2.0 larger than the simplistic Poisson errors.
We therefore quote error bars on all correlation functions of twice the 
Poisson error.}

\begin{figure*}
\begin{center}
\includegraphics[width=\textwidth]{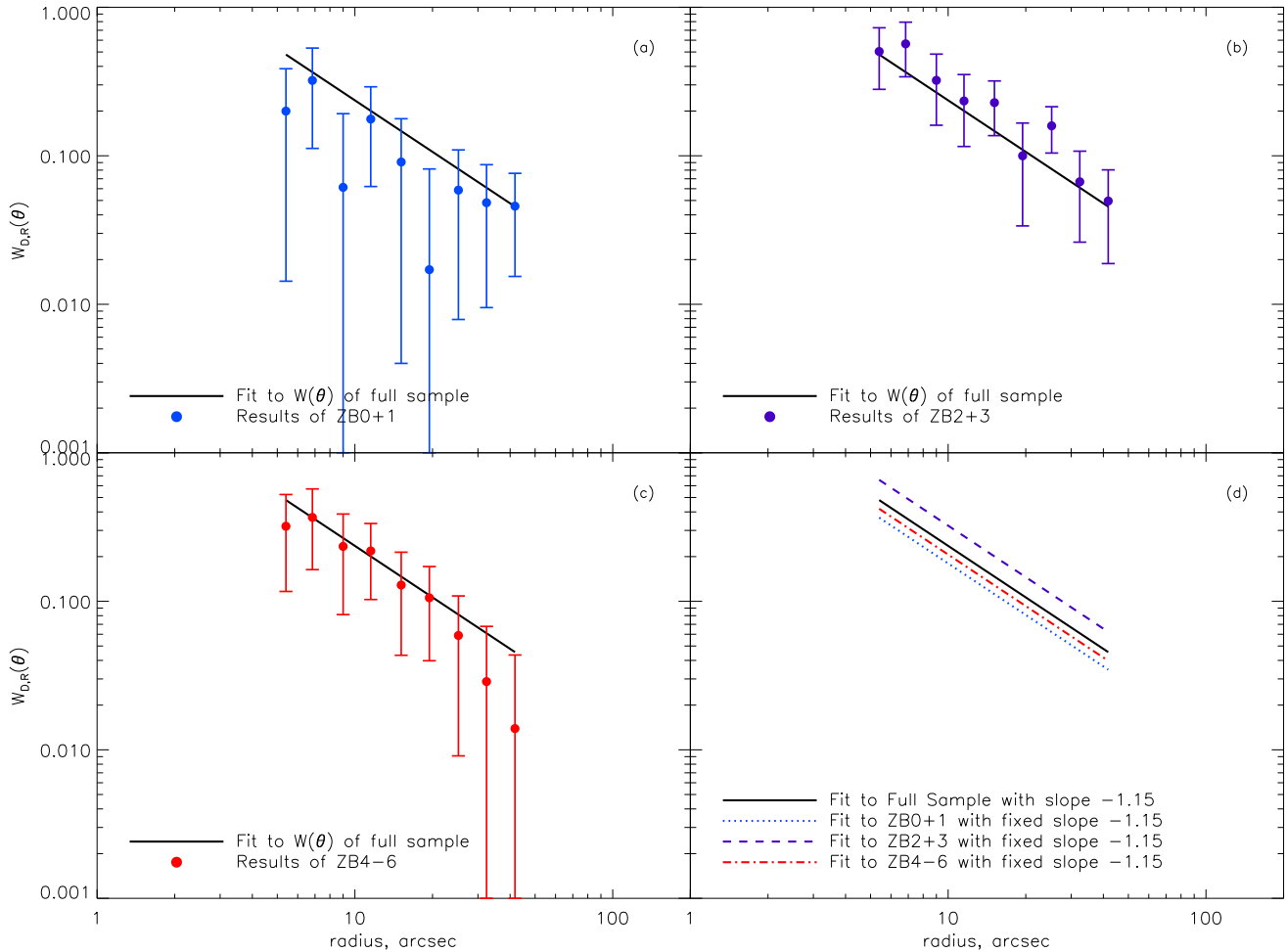}
\caption{Comparing cross-correlation functions of three redshift ranges ($D$) with the reference catalogue ($E$) 
being the full SIMPLE catalogue: (a) $W_{D,E}(\theta)$ where $D$ is the subset in $0.0\leq z<0.6$; (b) 
$W_{D,E}(\theta)$ where $D$ is the subset in $0.6\leq z<1.0$; (c) $W_{D,E}(\theta)$ where $D$ is the subset in 
$1.0\leq z<2.0$.  On each of (a)-(c) the black line shows the fit to the autocorrelation function of the full 
catalogue for comparison. {Error bars are twice the Poisson errors, as described in the text.}
The power-law fits to the four functions are shown in (d), where the slope has been fixed 
by the fit to the full catalogue.}
\label{fig:wtheta_z}
\end{center}
\end{figure*}

The strong clustering implies a significant correction to the measured
fluxes from stacking catalogue sources.  The correction accounts for
the flux contribution from any background sources separated by some
angular distance $\theta$ from the target.  This contribution, as a
fraction of the average source flux, is given by
Equation~\ref{eqn:acorrinteg}: a convolution of the correlation
function $W(\theta)$ with the beam profile for the corresponding band
(assumed to be Gaussian, $\exp(-\theta^2/2\sigma^2)$, with
$\sigma=$FWHM$/(2\sqrt{2\ln{2}})$), scaled by the number density of
background sources $n$.  

\begin{equation}
 F=n\int_0^\infty W(\theta) e^{-\theta^2/2\sigma^2} 2\pi \theta d\theta
\label{eqn:acorrinteg}
\end{equation}

This equation gives the average contribution
of confused sources to a measured flux, hence a correction factor of
$1/(1+F)$ must be applied to stacked fluxes.

For the 24-\mum\ case a slightly different convolution must be used
because aperture photometry is used.  The contribution of a background
source to a flux measurement then depends not only on where it falls
on the beam profile, but on how much of its beam falls within the
aperture.  We computed the convolution of the 24-\mum\ point-response
function (PRF)\footnote{We used the empirical xFLS PRF available on the SSC website:
\url{http://ssc.spitzer.caltech.edu/mips/psf.html}} with the 13-arcsec
radius aperture, to give a curve of growth which represents the
contribution of a background source to the aperture as a function of
angular separation $\theta$.  This function is then substituted for
the Gaussian beam profile in Equation~\ref{eqn:acorrinteg}.

This method corrects a stacked flux using the average probability of
confusion from another source at separation $\theta$, scaled by the
amount of flux expected from a distance $\theta$ from the centre of
the beam.  When correcting stacks of individual redshift bins, we must
assume the same level of clustering in each bin if we are to use the
autocorrelation of the full catalogue.  To account for the probability
of confusion of a target from a particular redshift range, while
accounting for the contribution from background sources at all
redshifts, we must consider the cross-correlation of the sources in
the particular range (the `data' centres, $D$) with the full catalogue
(the `reference' centres, $E$; see Fig.~\ref{fig:wtheta_z}).  A
modification of the \citet{landy_bias_1993} method was used to
calculate the cross-correlation function $W_{D,E}(\theta)$, given by
Equation~\ref{eqn:landy_szalay_x}.  

\begin{equation}
 W_{D,E}(\theta)=\dfrac{DE - 2\,DR - RR}{RR}
\label{eqn:landy_szalay_x}
\end{equation}

The robustness of the results was tested by checking against the
method of \citet{masjedi_very_2006}, which gave indistinguishable
results.

Thus we calculated the average across all the data centres, of the
excess probability of confusion with any of the reference centres.
The correction to stacked flux was then calculated in the same way as
described above, using $W_{D,E}$ in Equation~\ref{eqn:acorrinteg}.  It
should be noted that using this estimate of the fractional
contribution involves the implicit assumption that the average flux of
background sources is equal to the average stacked flux.  Since we can
only correct for confusion with catalogue sources by this method
(i.e. to avoid double-counting) this is a reasonable assumption.

We calculated the corrections for three redshift ranges, shown in
Table~\ref{tab:clustercorr}, by grouping the bins as follows: $0.0\leq
z<0.6$; $0.6\leq z<1.0$; $1.0\leq z<2.0$.  Errors in the table were
calculated using standard formulae for the propogation of errors, with
the error bars on $W_{D,E}(\theta)$ as shown in
Fig.\ref{fig:wtheta_z}.  The results in the table are not surprising:
160\mum\ has the lowest resolution therefore the greatest confusion,
24\mum\ suffers more than 70\mum\ because aperture photometry is used,
and the radio images have sufficiently high resolution to largely avoid confusion.

\begin{table*}
\begin{center}
\begin{tabular}{ l c c c c }
\hline
 & \multicolumn{3}{c}{$C$ (cross-correlation)} & $C$ (autocorrelation)\\ 
Band & $0.0\leq z<0.6$ & $0.6\leq z<1.0$ & $1.0\leq z<2.0$ & All \\ \hline
24\mum & $0.86\pm 0.05$ & $0.74\pm 0.04$ & $0.80\pm 0.04$ & $0.79\pm 0.03$ \\
70\mum & $0.90\pm 0.04$ & $0.80\pm 0.03$ & $0.86\pm 0.04$ & $0.84\pm 0.02$ \\
160\mum &$0.81\pm 0.05$ & $0.66\pm 0.04$ & $0.74\pm 0.05$ & $0.72\pm 0.03$ \\
1.4~GHz & -- & -- & -- & $1.000\pm 0.001$ \\
610~MHz & -- & -- & -- & $1.000\pm 0.002$ \\
\hline
\end{tabular}
\end{center}
\caption{Correction factors ($C$) to stacked source fluxes to remove contribution from correlated background sources, 
using autocorrelation of full catalogue and cross-correlations of three redshift ranges with the full catalogue.  
{Errors were calculated as described in the text.}  Corrected flux $S_\text{stack,corr}=S_\text{stack}\times C$}
\label{tab:clustercorr}
\end{table*}

\section{Galaxy SEDs and $K$--Corrections}
\label{sec:kcorrs}

\subsection{Radio $K$--Correction}

\begin{figure}
\begin{center}
\includegraphics[width=0.45\textwidth]{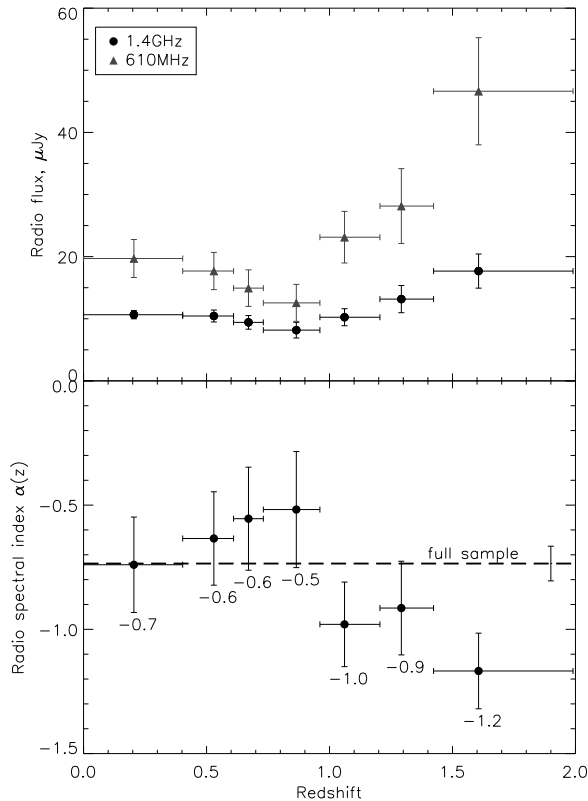}
\caption{Top: stacked radio fluxes {(and $1\sigma$ errors)}
as a function of redshift.  Note the increase in 610-MHz flux at high redshift 
which appears to be the main driver for the evolution in radio spectral index shown below.
Bottom: radio spectral index as a function of redshift, calculated
from fluxes shown above.  The dashed line (with error bar) represents
the spectral index calculated from stacks of the full sample,
$-0.74\pm 0.07$ (consistent with the mean of the indices of all the
bins).  The points represent the seven redshift bins and are labelled
with the corresponding values of $\alpha$, with {$1\sigma$} errors of $\pm 0.2$
calculated as described in Section~\ref{sec:errors}.
Horizontal error bars mark the widths of the bins.}
\label{fig:alpha}
\end{center}
\end{figure}

Observed fluxes were converted to rest-frame (emitted) monochromatic
luminosities using Equation~\ref{eqn:l_mon}, which contains a
bolometric $K$-correction $K(z)$, accounting for the shift of the
spectrum in relation to the receiver, and a further bandwidth
correction $[1+z]^{-1}$, accounting for the stretching of the spectrum
in relation to the bandwidth of the receiver ($d_L$ is the luminosity
distance to the source, while $z$ is its redshift).
\begin{equation}
L_{\nu,em}=4\pi d_L^2\ S_{\nu,\text{obs}}\ K(z)\ [1+z]^{-1}
\label{eqn:l_mon}
\end{equation}

The radio spectrum can be assumed to follow a simple power law
($S_\nu\propto\nu^\alpha$) resulting from the sum of the non-thermal
synchrotron and thermal bremsstrahlung components; the power law index
is typically $\alpha\approx-0.8$ for star-forming galaxies
\citep{condon_radio_1992}, although steeper indices might be expected 
in AGN--dominated sources \citep{ibar_deep_2009_ii}.  
The $K$--correction to a monochromatic flux
with a power law spectrum is given by Equation~\ref{eqn:radio_k},
which is independent of the filter transmission function.

\begin{equation}
K(z) = [1+z]^{-\alpha}	
\label{eqn:radio_k}
\end{equation}

The radio spectral index for each bin was evaluated using the stacked
fluxes in the two radio bands, $S_{\nu}^{\text{610~MHz}}$ and
$S_{\nu}^{\text{1.4~GHz}}$ in Equation~\ref{eqn:radio_alpha} (which
follows from $S_\nu\propto\nu^\alpha$), and is plotted in
Fig.~\ref{fig:alpha}.
\begin{equation}
\label{eqn:radio_alpha}
\alpha = \dfrac{\log{(S_{\nu}^{\text{610~MHz}}/S_{\nu}^{\text{1.4~GHz}})}}{\log{(610/1400)}}
\end{equation}

These spectral indices were used to $K$--correct each measured radio
flux using Equation~\ref{eqn:radio_k}, taking the observed median
index for each bin to calculate $K$--corrections for all sources in 
that bin.  

In Fig.~\ref{fig:alpha} we note an apparent
evolution to steeper radio slopes at increasing redshift in our
sample.  A linear least-squares fit to the $\alpha(z)$ values gives a
slope of $-0.39\pm0.15$; the slope is non-zero at the $2.6\sigma$
level.  This apparent trend is an unexpected result, and it is
noteworthy that it was not observed in the stacked 24-\mum\ sample of
\citet{ivison_blast:far-infrared/radio_2009}, who used the same radio
data and stacking technique; although their spectral
indices do cover a similar range.  The possible implications are
discussed in Section~\ref{sec:radio-disc} of this paper, but we note
that using a single spectral index of $-0.74$ for $K$--corrections leads
to a slight rise in the $q$ indices in the three highest redshift bins
(a change of $\delta q = +0.17$ for the last bin at $\langle z \rangle=1.6$).

\subsection{Infrared $K$--Correction}
\label{sec:seds}

\begin{figure}
\begin{center}
\includegraphics[width=0.45\textwidth]{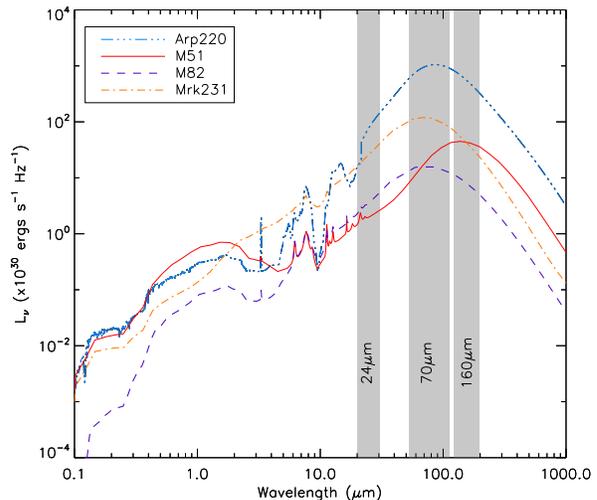}
\caption{Comparison of four SED templates showing relative positions of MIPS filters.  Note the differences in the 
shapes blueward of the filters, in particular the strength of the PAH and silicate features in Arp220, the power-law 
slope resulting from AGN-heated dust in Mrk231, and the cold-dust bump of M51 peaking at a longer wavelength.}
\label{fig:seds}
\end{center}
\end{figure}
\begin{figure}
\begin{center}
\includegraphics[width=0.4\textwidth]{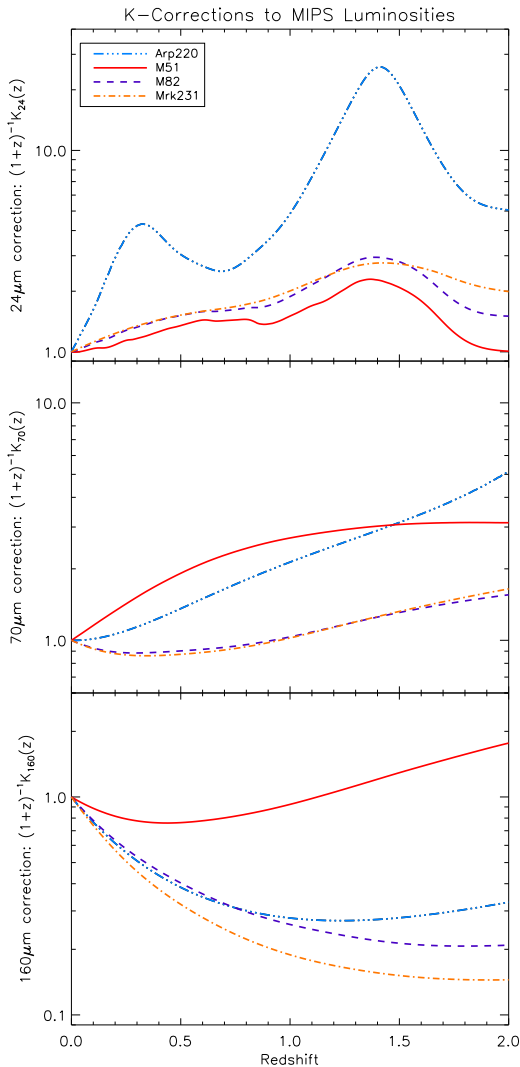}
\caption{$K$--corrections for the three MIPS bands (from top to bottom, 24\mum, 70\mum, 160\mum), calculated from 
four different SED templates.  Note the effect of the strong silicate troughs in Arp220 on the 24-\mum\ flux at 
redshifts 1.4 (rest-frame wavelength of 10\mum) and 0.3 (rest-frame 18\mum), and the effect of the different dust 
temperatures of the galaxies on the 70 and 160-\mum\ corrections.}
\label{fig:mipsk}
\end{center}
\end{figure}
In the mid-/far-infrared part of the spectrum sampled by the MIPS
bands, the assumption of a simple power law is not valid and
$K$--corrections must be calculated by evaluating
Equation~\ref{eqn:general_k}, which defines the $K$--correction as the
ratio of intrinsic luminosity to observed for a general filter
transmission profile $T_\nu(\nu)$.
\begin{equation}
K(z) = \dfrac{\int_0^\infty T_\nu(\nu) L_\nu(\nu) d\nu}{\int_0^\infty T_\nu(\nu) L_\nu(\nu[1+z]) d(\nu[1+z])}
\label{eqn:general_k}	
\end{equation}
This requires knowledge of both the filter transmission
function\footnote{The transmission functions for the MIPS filters are
available on the \textit{Spitzer} Science Center website at
\url{http://ssc.spitzer.caltech.edu/mips/spectral_response.html}}
$T_\nu(\nu)$ and the SED $L_\nu(\nu)$.  {A well-studied local
galaxy can be used} as a template for high-redshift galaxies; commonly used
templates in FIR studies include Arp220 and M82, which are IR-luminous
and therefore considered to be more typical of IR-selected
galaxies at high redshift.  However in the current study there is no
reason to presume that these dusty IR-luminous objects are
representative of the stacks, since our sample is selected in the IRAC
3.6 and 4.5-\mum\ channels.

We tried a range of different template SEDs to compare the results
given by the various $K$--corrections.  Four templates were chosen to 
represent different types of IR SEDs:
\begin{enumerate}
 \item Arp220 \citep[template from][]{pope_hubble_2006}: a bright ULIRG with a particularly large mass of hot dust and high star~formation following a recent merger
 \item M51 \citep[GRASIL template:][]{silva_modelingeffects_1998}: A typical large late-type spiral with moderate star~formation and cold dust distributed in the spiral arms
 \item M82 \citep[GRASIL template:][]{silva_modelingeffects_1998}: The prototype hot starburst galaxy, with intense star~formation probably triggered by a tidal interaction with M81.
 \item Mrk231 \citep[GRASIL template:][]{vega_modellingspectral_2008}: A Seyfert--1, probable merging system, with both starburst and AGN components
\end{enumerate}

The four SED templates are shown in Fig.~\ref{fig:seds}, and the
$K$--corrections for the three MIPS filters are plotted in
Fig.~\ref{fig:mipsk}.  The stacking was repeated four times,
using a different template for $K$--corrections each time, to compare 
the effects of different assumptions about the SEDs.  The results are 
described in Section~\ref{sec:kcorrs-disc}.

\section{Results and Discussion}
\label{sec:discussion}
\subsection{Evolution of Radio Properties of the Sample} 
\label{sec:radio-disc}
Our results indicate a significant increase in radio luminosity with
redshift (see Table~\ref{tab:sfr}), which is to be expected if the
radio emission is related to star formation, due to the increase in
star-formation activity in the most massive galaxies from the local
universe back to $z\sim2$.  The apparent evolution in radio spectral
index over the redshift range (Fig.~\ref{fig:alpha}) is more
surprising and, notwithstanding the large error bars, hints at a
fundamental change in the sample demographic, with different sources
dominating the radio luminosity at $z<1$ and $z>1$ respectively.  The
most likely potential contaminant is radio flux from AGN, which would
have a different spectral index than that from star formation, and
would also have the effect of boosting radio luminosity.  Radio-loud
AGN source counts are known to evolve strongly at $z>1$ 
\citep[e.g.][]{wall_parkes_2005}.

The effect that AGN contamination would have on the median spectral indices
is not entirely straightforward.  While flat ($\alpha\gtrsim-0.5$) 
spectra are associated with radio-quiet quasars or low-luminosity AGN 
\citep{bondi_vvds-vla_2007,huynh_radio_2007}, steep spectra
($\alpha <-1$) have frequently been used to select powerful radio galaxies at high 
redshift \citep[generally $z\gtrsim2$; e.g][]{de_breuck_sample_2000,pedani_efficiency_2003,cohen_deep_2004}.
This is because AGN radio spectra are flat at low frequencies but steepen at high 
frequency, hence steep slopes are observed when the spectrum is highly redshifted
(the frequency of the turnover varies, depending on properties such as magnetic field 
strength and electron density; \citealp{huynh_radio_2007}).  The evolving spectral
indices seen in Fig.~\ref{fig:alpha} could therefore be a sign of AGN dominating 
the radio signal at higher redshifts.

\begin{figure}
\begin{center}
\includegraphics[width=0.45\textwidth]{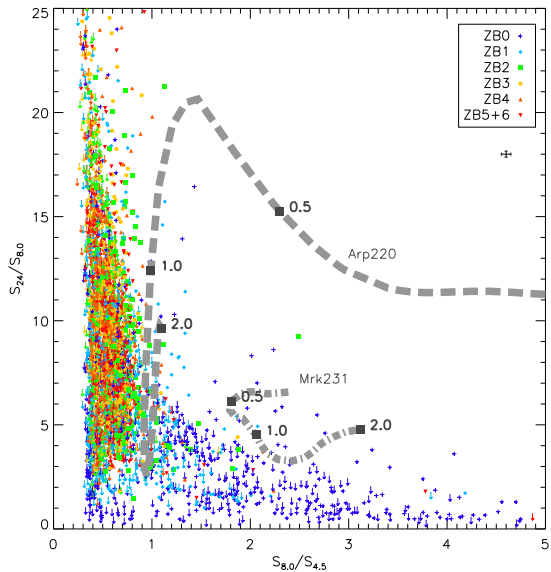} 
\caption{Observed-frame $S_{8.0\mum}/S_{4.5\mum}$ -- $S_{24\mum}/S_{8.0\mum}$ colour-colour plot of all objects in 
the sample, coloured by redshift bin (for ranges see Table~\ref{tab:zbins}).  
{Our 24-\mum\ aperture fluxes are used where they are $\geq 5\sigma$ (roughly 30\% of the positions).  
For the positions without a $5\sigma$ detection at 24\mum, we use the $5\sigma$ upper limit (indicated by arrows).
IRAC fluxes are from the SIMPLE catalogue.  
The representative error bar in the top right shows the median $1\sigma$ errors.  Some scatter in the vertical axis
is introduced by confusion at 24\mum, but positions in the horizontal axis are reliable as signal-to-noise 
is good in both 4.5\mum\ and 8.0\mum\ and confusion noise is much lower.}  Tracks of Arp220 and Mrk231 are 
overlaid in grey, showing the locus of each on the diagram as a function of redshift between 0 and 2.  This plot can 
be used as an AGN/starburst diagnostic since AGN have been shown to lie to the right, with 
$S_{8.0\mum}/S_{4.5\mum}>2$ at intermediate to high redshifts (as the prototype Mrk231 does) while starbursts (such 
as Arp220) lie above and to the left \citep[e.g.][]{ivison_spitzer_2004, pope_mid-infrared_2008, 
coppin_mid-infrared_2010}.  With the exception of bin ZB0 ($0.0\leq z<0.4$), our data lie well to the left of the 
plot, indicating flat spectral slopes at $<8.0\mum$\ (compared with Arp220) and negligible AGN contamination.  The 
scatter in bin ZB0 is attributed to strong PAH emission at 7.7\mum\ which makes the diagnostic unreliable at low 
redshift (although it is noted that strong PAH emission is generally associated with star formation and not AGN).}
\label{fig:irac}
\end{center}
\end{figure}

Matches with the Chandra X-ray catalogue \citep{virani_extended_2006}
have been removed from the sample, reducing the likelihood of
contamination from unobscured AGN.  
But overlap between X-ray and radio AGN is known to be small
\citep[e.g][]{pierce_effects_2010}.
In order to identify any AGN that are obscured
or undetected in X-rays, we looked to the MIR fluxes from the IRAC
catalogue.  IRAC colours have been shown to provide some limited
diagnostics for selecting obscured AGN based on the rest-frame MIR
slope \citep[e.g.][]{lacy_obscured_2004, stern_mid-infrared_2005,
alonso-herrero_infrared_2006, donley_spitzer_2007,
donley_spitzers_2008}.  In Fig.~\ref{fig:irac} we plot
$S_{8.0\mum}/S_{4.5\mum}$ against $S_{24\mum}/S_{8.0\mum}$ for all the
objects in the sample.  In this plot, the majority of objects have
$S_{8.0\mum}/S_{4.5\mum}<1$, well separated from the region occupied
by AGN at intermediate to high redshifts
\citep[e.g.][]{ivison_spitzer_2004, lacy_obscured_2004}.  
The scatter of colours in the lowest-redshift bin is due
to the presence of the 7.7-\mum\ PAH feature in the sources, and the
diagnostic is not reliable for this bin.  However, the locus of
objects in all other bins on the diagram clearly indicates that AGN
contamination is negligible. 

Taking Fig.~\ref{fig:irac} as evidence against significant AGN contamination,
we deduce that the radio emission originates from star-formation
activity, and that the index of the non-thermal continuum from CR
electrons evolves.  The model described by \citet{lacki_physics_2009-1} 
and \citet{lacki_physics_2009} {predicts
steep radio spectra as a result of increasing CR electron losses via
inverse-Compton scattering with the cosmic microwave background.  
That model predicts these losses to become significant in normal galaxies 
at $z\approx 2$, which seems to be supported by our data.  
Alternatively, the steepening spectral slope can be a sign of increased electron
calorimetry, as electron escape becomes less important relative to
electron cooling (including synchrotron, bremsstrahlung and inverse-Compton
losses; \citealt{lacki_physics_2009}).
}

It is possible that the evolution we see in the radio spectral index
is not a variation between galaxies at different redshifts, 
but a function of the rest-frame frequency that is observed, 
and that the assumption of a single power-law spectrum is flawed.  A curved spectrum
with power-law index increasing with radio frequency would produce a similar effect
when viewed at successively higher redshifts.  The \citet{lacki_physics_2009-1} 
models predict a steepening of the spectral index by only $\sim0.05-0.1$ dex (depending 
on gas surface density) between the frequencies probed by 1.4-GHz observations at redshifts
from 0 to 2.  The evolutionary fit to our data indicates a change of $-0.8\pm0.3$ over this
range, suggesting that actual evolution with redshift does occur in the sample. 

To better understand the change in the sample demographic across the
redshift bins, we must consider the distribution of stellar masses in
the respective bins.  Malmquist bias means that our sample is
increasingly dominated by the most massive galaxies towards higher
redshifts.  Stacking a sub-sample limited to masses
$>10^{11}\text{M}_{\odot}$ is unhelpful due to small-number statistics, but
repeating the stacking analysis with a mass limit of $\log(M_\star)
\geq 10.5$ gave results for both $\alpha$ and $q$ that were fully
consistent with the full sample, although error bars were large (the
lowest-redshift bin contained too few objects to obtain a reliable value of
the radio spectral index).  Results can be seen in Fig.~\ref{fig:q_tir}.

\subsection{The Observed and $K$--corrected FIR--Radio Correlation as a Function of Redshift} 
\label{sec:kcorrs-disc}

\begin{figure*}
\begin{center}
\includegraphics[width=\textwidth]{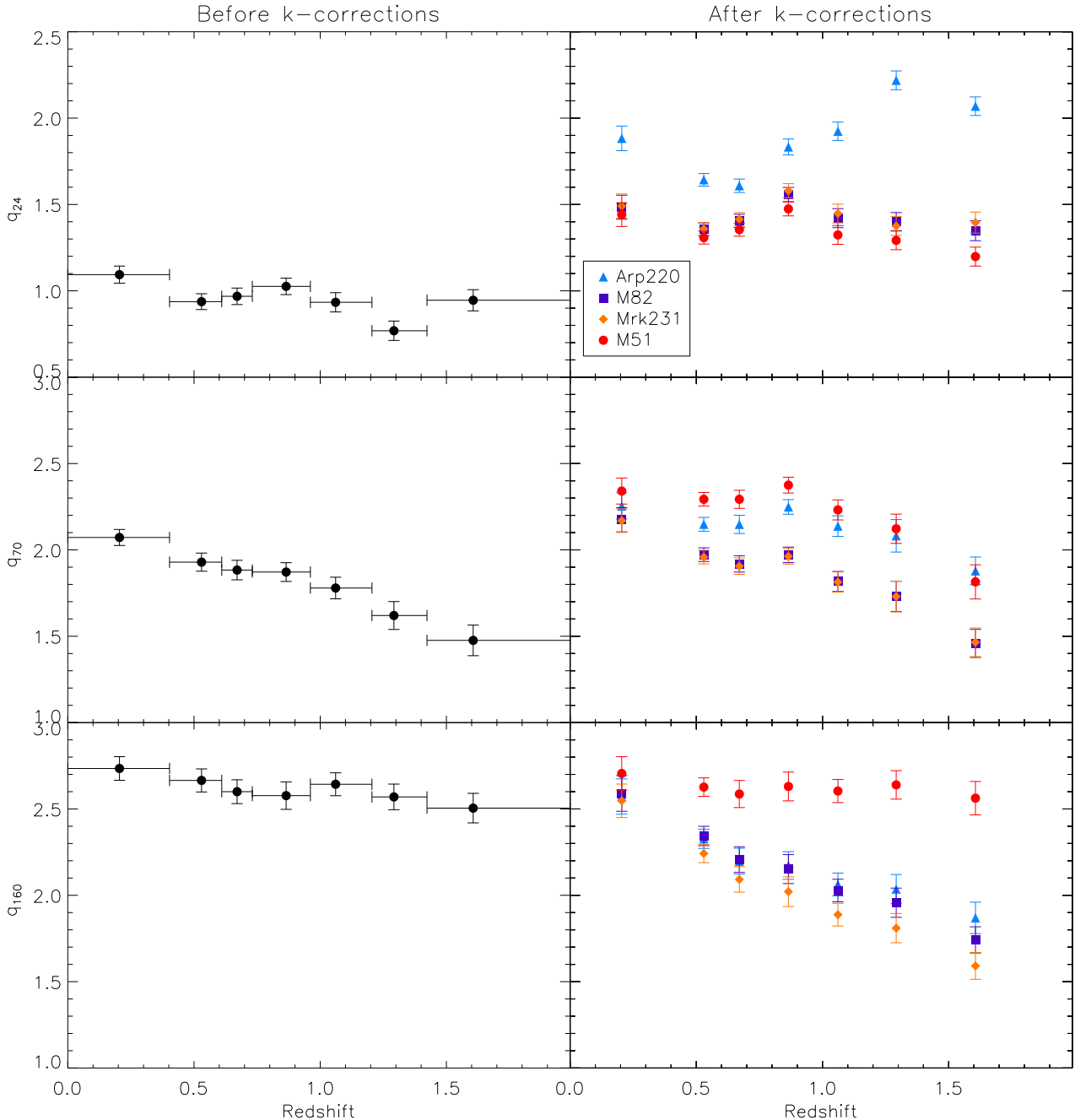} 
\caption{Far-infrared--radio relation ($q$) as a function of redshift for the three MIPS bands (from top to bottom, 
24\mum, 70\mum, 160\mum).  Ratios of stacked observed fluxes are plotted on the left.  On the right we show ratios of 
stacked $K$--corrected fluxes: radio fluxes are $K$--corrected using the measured spectral index for each bin, and 
infrared fluxes are $K$--corrected using the four templates described in Section~\ref{sec:seds}.  Vertical error bars 
represent estimated $1\sigma$ uncertainties on the stacked medians, following \citet{gott_median_2001}: they indicate 
the spread of the data, and are of similar size or larger than the noise, as described in Section~\ref{sec:errors}.  
Horizontal bars in the left-hand panels mark the full width of each bin.  Horizontal bars are omitted from the 
right-hand panels for clarity.}
\label{fig:q_z}
\end{center}
\end{figure*}

The $q$ index was calculated for each stack and for each FIR band
$i$, substituting the monochromatic FIR flux $S_i$ (Jy) for $S_\text{IR}$
in Equation~\ref{eqn:q_ir}.  Details of the stacked fluxes and $q$
ratios can be found in Table~\ref{tab:stackresults} at the end of this
paper.  Fig.~\ref{fig:q_z} displays the calculated $q$ values as a
function of redshift, both before and after $K$--corrections were applied.
Fluxes were $K$--corrected individually, rather than after stacking, using
the photometric redshift of each source, and for the radio, the stacked 
spectral index measured in the corresponding bin, and for the FIR, the 
corrections shown in Fig.~\ref{fig:mipsk}.  

The left panel of Fig.~\ref{fig:q_z} reveals a very slight downward trend of $q$
in each of the three MIPS bands, with a more bumpy evolution in $q_{24}$.  
The anomalies in the observed $q_{24}$ evolution can be explained in terms of the
MIR SED, which in star-forming galaxies often contains broad PAH
emission features at 3.3, 6.2, 7.7, 8.6, 11.3 and 12.7\mum\
\citep[e.g.][]{roche_atlas_1991, genzel_what_1998,
armus_observations_2007}.  These redshifted features can account for the boosting
of $q_{24}$ in the bins centred at redshifts 0.9 and 1.6. Similarly the dip in
$q_{24}$ at redshift 1.3 can be attributed to the
broad 10-\mum\ silicate trough redshifted into the 24-\mum\ passband.
The width of the redshift bins and the use of photometric redshifts
accounts for the breadth of redshifts over which these features appear
to have an effect.\footnote{The apparent evolution of $q$ with redshift that is expected from various SED
templates has been plotted over observed data by several authors including \citet{ibar_exploringinfrared/radio_2008},
\citet{seymour_investigatingfar-ir/radio_2009} and \citet{sargent_vla-cosmos_2010}.
In this work we choose instead to plot the $K$--corrected $q$-values in an attempt to
emphasise the `excess' evolution that may reveal intrinsic changes in the rest-frame flux ratios
of galaxies within the sample at different redshifts.}

The right panel of Fig.~\ref{fig:q_z} shows the effects of $K$--correcting
MIPS fluxes using the four SED templates introduced in
Section~\ref{sec:seds}.  Overall, the M51 template gives rise to the least 
evolution in all three $q$ indices.  $K$--corrections using the Arp220 template
give rise to an increasing $q_{24}$, due to the steeper MIR slope
($\lambda \lesssim 24\mum$).  In $q_{160}$ the Arp220, Mrk231 and
M82 templates all exacerbate the downward trend towards high redshift, 
while the M51 template removes it, as a result of the cooler dust temperature (longer
wavelength of the peak) in M51.  However, none of the templates removes the trend in 
$q_{70}$, and this could be attributed to a real evolution in rest-frame flux ratios, 
or a steeper spectral slope at $\lambda \lesssim 70\mum$ in the galaxies sampled, 
in comparison to the templates chosen.

\subsection{Infrared Spectral Energy Distributions}
\begin{figure}
\begin{center}
\includegraphics[width=0.45\textwidth]{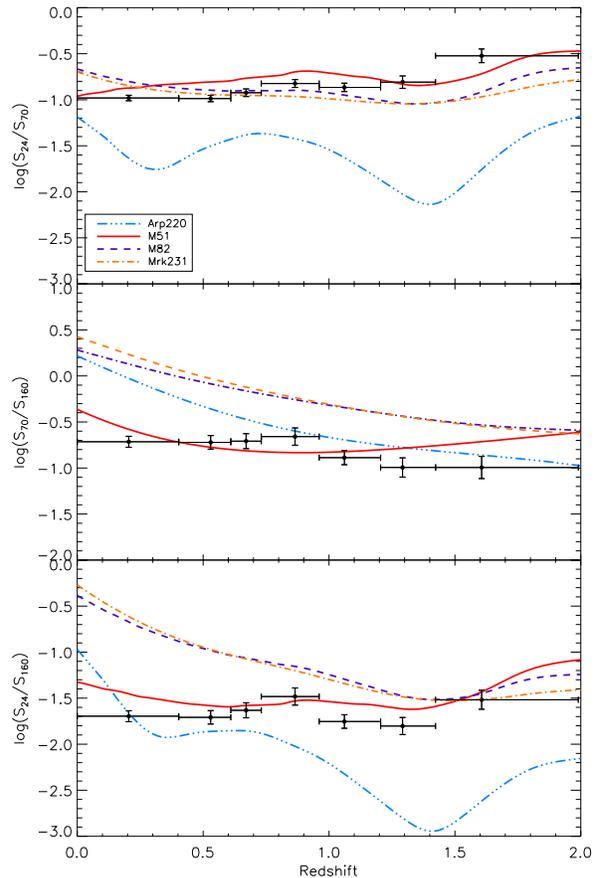} 
\caption{Observed MIPS flux ratios as a function of redshift for the three MIPS bands.  From top to bottom: 
$\log(S_{24}/S_{70})$, $\log(S_{70}/S_{160})$, $\log(S_{24}/S_{160})$.  Overlaid are the tracks of the four templates 
described in Section~\ref{sec:seds}.  Error bars are as in Fig.~\ref{fig:q_z}.}
\label{fig:colors}
\end{center}
\end{figure}

Interpretation of our results is evidently subject to the assumptions
made about the `average' or typical SED of the sample.  It is possible
to better constrain the FIR $K$--correction by analysing the evolution
of MIPS flux ratios (colours) as a function of redshift.  These
colours are sensitive to the position of the peak of the thermal dust
emission, hence the temperature of the emitting dust, as well as the
slope of the SED on the short-wavelength side of the peak.
Fig.~\ref{fig:colors} shows the evolution in observed MIPS colours
with redshift, plotted over the expected tracks for each of the SED
templates, and reveals that the SED most consistent with observed
colours at all redshifts is M51.

The important factor distinguishing the M51 template from the others
used is the position of the peak of the SED at a longer wavelength.
M51 is a quiescent star-forming galaxy with an IR SED dominated by
cold dust, and evidence from Fig.~\ref{fig:colors} therefore points to
a cold dust temperature for the galaxies in our sample, at least in
the first three redshift bins.  The 70--160-\mum\ colour is directly
sensitive to the position of the peak at low redshifts, but it is
clear from the middle panel of Fig.~\ref{fig:colors} that over the
last four bins the M51 and Arp220 templates are barely distinguishable
in this colour space, so we cannot draw conclusions on the dust
temperatures at $z\gtrsim 0.8$.  This is because at these redshifts
both bands are shortward of the peak of even the hottest IR SED, and
probe the slope on the Wien side. The 70--160-\mum\ colours of the
high-redshift bins are consistent with the steeper slopes of M51 and
Arp220, and not with the shallower slopes of M82 and Mrk231 (similarly
the 24--70-\mum\ and 24--160-\mum\ colours rule out Arp220, due to its
strong PAH emission).  These steeper slopes are potentially an
indication of a stronger contribution from `cold' dust (in the ambient
ISM) relative to `hot' dust (in H\textsc{ii} regions associated with
star formation) or a dearth of emission from very small grains (VSGs); alternatively
they could even result from extremely optically thick systems where
the SED is steepened by MIR dust attenuation. In this case, however,
we would expect to see a stronger 10-\mum\ silicate absorption feature
such as that evident in the Arp220 $K$--correction at z$\sim 1.5$. Our
stacked colours are not consistent with such a strong absorption which
reduces the likelihood that optically thick MIR emission is
responsible for the steeper rest-frame MIR slope at high redshift.

Cold dust temperatures are nevertheless consistent with the
conclusions of studies such as \citet{chapman_redshift_2005} and
\citet[2008]{pope_hubble_2006} for high-redshift sub-mm galaxies
(SMGs), and \citet{symeonidis_link_2009}, {\citet{seymour_comoving_2010}
and \citet{giovannoli_population_2010} for 70-\mum-selected galaxies
at $z\lesssim1$.}  There are also parallels with a recent detailed study of
two massive $K$--selected galaxies at $z\sim2$ by
\citet{muzzin_well-sampled_2010}, who fitted SEDs to data from
\textit{Spitzer}, BLAST and LABOCA \citep{siringo_large_2009} instruments.  Their
best fits were star-formation-dominated SEDs with
$L_\text{TIR}\sim10^{13}\text{L}_{\odot}$, but with cold dust temperatures, in
contrast to ULIRGs in the local universe. Similar cool SEDs have also
been determined for SMGs out to $z\sim 1$ from
BLAST and \textit{Herschel} studies \citep{dye_radio_2009,amblard_herschel_2010}.

For our sample of massive galaxies, we expect to probe the epoch of
stellar mass buildup at $z>1$.  Indeed, Table~\ref{tab:sfr} shows that
both radio-- and IR--derived SFRs in our bins do reach high values
beyond this redshift.  It seems a reasonable assumption that the IR
and radio luminosities are dominated by star-forming activity, since
we do not expect a significant contamination from AGN-dominated
galaxies in the sample (see Section~\ref{sec:radio-disc}).  We observe 
therefore that despite the tendency towards higher
luminosities (and SFRs) in the sample at increasing redshifts, there
is no evidence for a change in the SED towards the templates of local 
high-SFR galaxies such as Arp220 or M82.

\begin{table*}
\begin{center}
\begin{tabular}{ c c c c c c c }
\hline
$z$ Range  & $\langle z\rangle$ & $L_\text{1.4 GHz},\ \text{W Hz}^{-1}$  & $\text{SFR}_\text{1.4 GHz},\ \text{M}_{\odot}\text{yr}^{-1}$ & $L_\text{TIR},\ \text{L}_\odot$ & $\text{SFR}_\text{TIR},\ \text{M}_{\odot}\text{yr}^{-1}$& $q_\text{TIR}$ \\  \hline
0.00--0.40 & 0.21               & $ (1.03 \pm  0.13) \times 10^{21} $    & {$0.57 \pm 0.08$} & $(5.8  \pm 1.4)\times 10^{9} $ & {$0.61\pm 0.17$} & $2.76 \pm 0.12$  \\ 
0.40--0.61 & 0.53               & $ (1.21 \pm  0.06) \times 10^{22} $    & {$4.23 \pm 0.21$} & $(5.7  \pm 1.4)\times 10^{10}$ & {$4.8 \pm  1.3$} & $2.68 \pm 0.11$  \\
0.61--0.73 & 0.67               & $ (2.15 \pm  0.11) \times 10^{22} $    & {$7.50 \pm 0.37$} & $(9.9  \pm 2.5)\times 10^{10}$ & {$8.1 \pm  2.1$} & $2.67 \pm 0.11$  \\
0.73--0.96 & 0.87               & $ (3.15 \pm  0.19) \times 10^{22} $    & {$10.98\pm 0.69$} & $(1.72 \pm0.43)\times 10^{11}$ & {$18.4 \pm 4.7$} & $2.75 \pm 0.11$  \\
0.96--1.20 & 1.06               & $ (7.48 \pm  0.72) \times 10^{22} $    & {$26.1 \pm 2.5 $} & $(3.27 \pm0.82)\times 10^{11}$ & {$34.1 \pm 8.8$} & $2.65 \pm 0.12$  \\
1.20--1.42 & 1.29               & $ (1.52 \pm  0.14) \times 10^{23} $    & {$53.0 \pm 4.9 $} & $(6.2  \pm 1.6)\times 10^{11}$ & {$64   \pm 16 $} & $2.62 \pm 0.12$  \\
1.42--2.00 & 1.61               & $ (3.63 \pm  0.35) \times 10^{23} $    & {$127  \pm 12  $} & $(1.07 \pm0.27)\times 10^{12}$ & {$109  \pm 28 $} & $2.48 \pm 0.12$  \\
\hline
\end{tabular}
\end{center}
\caption{Stacked galaxy properties derived from measured fluxes:  Rest-frame 1.4-GHz luminosities; SFRs derived from 
$L_\text{1.4~GHz}$ using the \citet{bell_estimating_2003} calibration; Total IR luminosities ($L_\text{TIR}= 
L_{8-1000\mum}$) estimated directly from the $K$--corrected MIPS fluxes using the M51 template; SFRs derived from 
$L_\text{TIR}$ using the \citet{bell_estimating_2003} calibration; Corresponding $q$ values calculated as described 
in the text.  {Errors on $L_\text{1.4~GHz}$ are directly translated from the $1\sigma$ flux errors.}
Errors on $L_\text{TIR}$ are assumed to be 25\% as described in the text, while errors on SFRs are 
directly translated from luminosity errors, and do not include any systematics from the conversion to SFR.
{All SFRs are calibrated to a Kroupa IMF.}}
\label{tab:sfr}
\label{tab:lum}
\end{table*}

In Table~\ref{tab:sfr} we show indicative TIR luminosities, derived for 
each bin using the rest-frame FIR luminosities in the MIPS bands, scaled up
to $L_{8-1000\mum}$ assuming the M51 template.   {This was done by
using the luminosities in the three MIPS bands simultaneously to find the 
best-fitting normalization of the M51 template.} 
There will be some systematic uncertainties in the calibration,
and for an idea of the size of these we consider another method to estimate 
$L_\text{TIR}$ from MIPS luminosities.  \Citet{dale_infrared_2002} offer one 
such formula for $L_{3-1100\mum}$, calibrated for a large sample of normal 
star-forming galaxies with a range of morphologies, colours and FIR luminosities 
\citep[see also][]{dale_infrared_2001}; we therefore consider it appropriate
for M51--like SEDs. The uncertainty on this calibration was shown to be 
$\sim 25\%$ by \citet{draine_infrared_2007}, and we find that using 
Dale \& Helou's method yields values well within 25\% (typically 6\%, but as much 
as 16\% for the highest-redshift bin) of those found using the M51 template over the same range.  
Hence assuming systematic errors of 25\% on $L_\text{TIR}$ is reasonable.

Notwithstanding these uncertainties, the results imply that the typical
galaxies sampled have quiescent IR SEDs at low $z$, but rapidly evolve towards 
higher IR luminosities at increasing $z$.  By $z\sim2$ they appear to reach 
ULIRG luminosities, as star-formation activity becomes significantly more prevalent 
in massive galaxies at these redshifts \citep[e.g.][]{daddi_population_2005}. 
The rise in luminosity with redshift which we observe (from
$\sim10^{10}$ to $\sim10^{12}\text{L}_\odot$) may be partially attributed to
increasing median stellar mass with redshift.  This cannot be the full
story though, since assuming a linear relationship between stellar
mass and $L_\text{TIR}$ implies an increase by a factor of 19, whereas
$L_\text{TIR}$ increases by a factor of $\sim 180$, and $L_\text{1.4
GHz}$ by $\sim 220$ over the redshift range.
{Indeed we know that $L_\text{TIR}$ is linked not to stellar mass itself,
but to SFR, which is well-known to rise with increasing redshift 
\citep{lilly_canada-france_1996,madau_high-redshift_1996,
prez-gonzlez_stellar_2008,damen_evolution_2009,magnelli_0.4_2009}.}

\subsection{Evolution in Specific Star Formation Rates}
\label{sec:sfrs}
The SFRs given in Table~\ref{tab:sfr} were calculated using the formulae of \citet{bell_estimating_2003},
{which assume a \citet{salpeter_luminosity_1955} IMF}:
\begin{equation}
\text{SFR}_\text{TIR}  = \left\{
\begin{array}{r}   
    1.57\times10^{-10} L_\text{TIR}\left(1+\sqrt{\dfrac{10^9}{L_\text{TIR}}}\right),  \\
      L_\text{TIR}>10^{11}\\
    1.17\times10^{-10} L_\text{TIR}\left(1+\sqrt{\dfrac{10^9}{L_\text{TIR}}}\right),  \\
      L_\text{TIR}\leq10^{11}\\
\end{array}\right.  
\label{eqn:sfr_tir}
\end{equation}
\begin{equation}
\text{SFR}_\text{1.4 GHz}  = \left\{
\begin{array}{c r}   
    5.52\times10^{-22} L_\text{1.4 GHz},& L>L_c \\
    \dfrac{5.52\times10^{-22}}{0.1+0.9(L/L_c)^{0.3}}L_\text{1.4 GHz},& L\leq L_c \\
\end{array}\right. 
 \label{eqn:sfr_rad}
\end{equation}
where SFR is in units of $\text{M}_{\odot} \text{yr}^{-1}$, $L_\text{TIR}$ is given in units of 
$\text{L}_{\odot}$, $L_{\text{1.4 GHz}}$ in $\text{W Hz}^{-1}$ and $L_c=6.4\times10^{21}\text{W Hz}^{-1}$.  
These conversions were applied to the stacked $L_\text{TIR}$ and $L_{\text{1.4 GHz}}$ to obtain SFRs
and to stacked $L_\text{TIR}/M_\star$ and $L_{\text{1.4 GHz}}/M_\star$ 
to obtain SSFRs (where $M_\star$ are the individual stellar masses).
{All SFRs were converted to a \citet{kroupa_variation_2001} IMF by subtracting 0.2~dex, following 
\citet{damen_evolution_2009}, in order to ensure consistency with the stellar masses used.}
Radio- and TIR-derived SFRs appear to be roughly in agreement; the TIR values 
are generally higher although mostly they are within the broad error bars given
by the calibration of $L_\text{TIR}$.  Agreement naturally depends upon the value of $q_\text{TIR}$
as a function of redshift being equal to the local value (e.g. the median in Bell's (2003)
sample was 2.64).  This will be discussed in the next section.  We note that using a
constant radio $K$--correction based on the overall median spectral index of $-0.74$
reduces radio SFRs in the last three bins but does not improve the agreement overall.

\begin{figure}
\begin{center}
\includegraphics[width=0.49\textwidth]{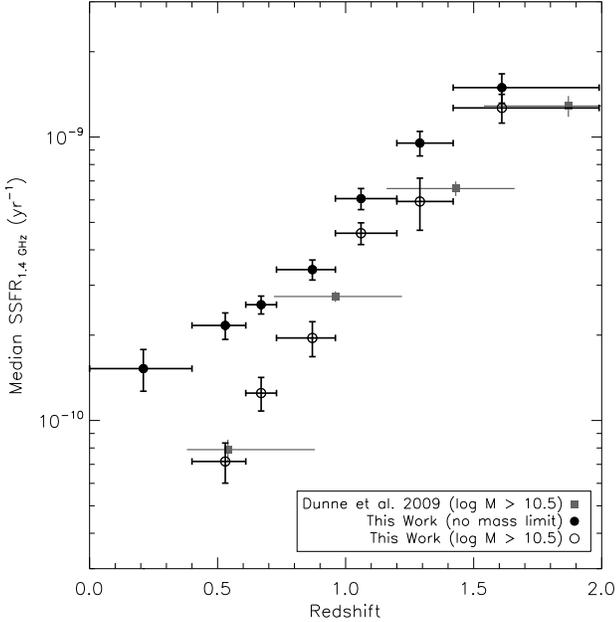} 
\caption{Medians specific SFRs from the 1.4~GHz luminosities as a function of redshift.
Solid black circles are the medians of the full stacked sample; open circles are
using the mass limit $\log{M_\star}\geq10.5$.  Also plotted are the results of 
\citet{dunne_star_2009} with the mass limit $\log{M_\star}\geq10.5$ (small {grey} symbols).  
All vertical error bars denote the estimated {$1\sigma$} uncertainty in the median following 
\citet{gott_median_2001}; systematics in the SFR calibrations are not included.  
Horizontal bars mark the widths of the bins.
{All SFRs in this plot have been converted to a Kroupa IMF as described in the text.}}
\label{fig:ssfrs}
\end{center}
\end{figure} 

Fig.~\ref{fig:ssfrs} (black solid points) shows the median specific SFRs (SSFRs: calculated 
source-by-source as the radio SFR divided by the stellar mass) 
as a function of redshift.  It is immediately apparent that
median SSFRs increase strongly with redshift, indicating a rise in
star-formation efficiency within the sample at increasing look-back
times, a result seen many times in the literature
\citep[e.g.][]{cowie_new_1996,madau_star_1998,brinchmann_mass_2000,
bauer_specific_2005,feulner_connection_2005,prez-gonzlez_stellar_2008,
dunne_star_2009,damen_evolution_2009,oliver_specific_2010}.
The black open points in Fig.~\ref{fig:ssfrs} show the results of the 
mass-limited sub-sample ($\log(M_\star)\geq 10.5$) in comparison to the 
full sample (filled points).  This shows the effect of having lower median 
stellar masses in the full sample at low redshifts in particular
(because SSFR is a function of stellar mass as well as redshift).
The consistency between the results of this stacking study with the
stacked $K$-selected sample of \citet{dunne_star_2009} 
seems to support the idea that the IRAC selection targets a similar 
population to $K$ selection.

\subsection{Does the FIR--Radio Correlation Evolve with Redshift?}
The M51--corrected $q$ indices in Fig.~\ref{fig:q_z}, which we have
shown to be the most appropriate, appear to show the least evolution
in the FRC.  It is important to note that the observed fluxes in the
three MIPS bands trace different parts of the SED and are affected by
different components of emission in the source galaxies.  In
particular, the observed 160-\mum\ flux between redshifts of 0 -- 2 is
the closest tracer of the FIR peak of the SED (due to emission from large
graphite and silicate grains, making up the majority of the dust
mass), and $q_{160}$ $K$--corrected with M51 displays no evidence of evolution.  
The $q_{24}$ index is also broadly constant, despite 24\mum\ being a closer tracer of
emission from PAHs than the bulk of the dust \citep[e.g.][]{desert_interstellar_1990}.
This lack of evolution is in agreement with previous studies of 24-\mum-selected
\citep[e.g.][]{appleton_far-_2004} and radio-selected samples
\citep[e.g.][]{ibar_exploringinfrared/radio_2008}.  Somewhat surprisingly however, 
some evolution is still apparent in $q_{70}$ after M51 $K$--correction, at around 
$3\sigma$ significance.  

Some of the anomalies in the $K$--corrected $q_{24}$ and $q_{70}$ graphs could be due to the MIR 
spectrum and/or the radio $K$--correction.  The $z\approx0.9$ bin for example coincides with 
the redshifted PAH feature at 12.7\mum, and the boost in $K$--corrected 24-\mum\ flux
at this redshift might be a sign of strong PAH emission in the sources.  
The radio $K$--correction could also play a part, since in this bin
the measured spectral index is relatively flat.  This explanation appears likely 
since a similar bump is apparent in $q_{70}$ at the same redshift. Furthermore, there
are particularly low values of $q_{24}$ and $q_{70}$ in the $z\approx1.6$ bin, which
coincides with the steepest measured spectral index.
Repeating the stacking analysis using a constant spectral index of $-0.74\pm 0.07$ for 
radio $K$--corrections was found to have a small effect on both of these bins, changing each 
$q$ index by $-0.06$~dex at $z\approx0.9$ and $+0.17$~dex at $z\approx1.6$.  
Similarly the values in the intermediate bins at $z\approx1.1$ and $1.3$ 
were raised by 0.07 and 0.06 respectively (changes in the low-redshift bins were 
negligible); however this still leaves a decline of $2\sigma$ significance in $q_{70}$ 
when the M51 template is used.  Clearly it is the FIR SED which dominates the evolution
of monochromatic $q$ indices, and not the radio spectrum.

One factor that could account for this decline in $q_{70}$ is a
steepening of the continuum slope shortward of 70\mum, relative to the
M51 template.  The SED in the MIR region ($10\mum \lesssim \lambda
\lesssim 70\mum$) is thought to be dominated by emission from very
small grains (VSGs, with radii $\lesssim 10$nm) with fluctuating
temperatures resulting from a mixture of thermal and single-photon
heating \citep{desert_interstellar_1990}.  A steepening of the the
slope shortward of 70\mum\ might be due to an increase in the FIR
($\sim 100$\mum) luminosity (dominated by big grains) relative to the 
VSG contribution at shorter wavelengths,
although this is not clear from the MIPS flux ratios (Fig.~\ref{fig:colors}).

In this context it is interesting to compare with the results of
\citet{seymour_investigatingfar-ir/radio_2009}, who measured 70-\mum\
fluxes for a sample of faint radio sources and reported a decrease in
observed $q_{70}$ with redshift (both for detected sources and
stacks), which is not fully accounted for by the $K$--correction of
any single model SED.
Seymour et al. concluded that their stacked data show a discrepancy at
$0.5\lesssim z\lesssim1.5$ between increasing total $L_\text{TIR}$
values (estimated from radio luminosities) and decreasing $q_{70}$,
implying a change in the ULIRG SED at high redshift.  Whatever the
cause, it seems plausible that these two samples are similarly
affected.

Some of the first results from \textit{Herschel} provide further tantalising 
evidence for some change in star-formation activity at high redshifts:
\citet{rodighiero_first_2010} stacked into 100 and 160-\mum\ imaging from the PACS 
\citep[][]{poglitsch_photodetector_2010} Evolutionary Probe \citep[PEP;][]{berta_dissecting_2010},
at the positions of IRAC (4.5-\mum) sources that were optically classified as star-forming and undetected in the 
160-\mum\ image, divided into bins of stellar mass and redshift.  They found that SSFRs (derived from IR+UV 
luminosities) followed a power-law trend with mass, with an index of $-0.25^{+0.11}_{-0.14}$ at 
$z<1$, in agreement with SSFRs from radio stacking \citep{dunne_star_2009,pannella_star_2009}, but
that the index steepened to $-0.50^{+0.13}_{-0.16}$ at $1<z<2$, deviating from the radio results.  
A change in the IR SED or $q_\text{TIR}$ would be expected to produce such a deviation between
SSFRs derived respectively from IR and radio (as is suggested by our data in Table~\ref{tab:sfr}).

\begin{figure}
\begin{center}
\includegraphics[width=0.5\textwidth]{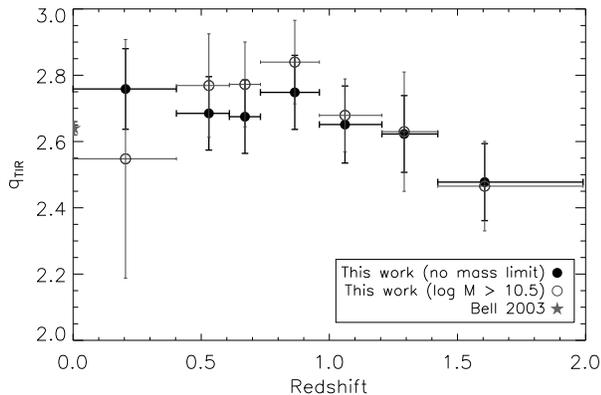} 
\caption{Total infrared luminosity--to--1.4~GHz luminosity ratio as a function of redshift for the full sample (black 
solid points) and for the mass limited sample ($\log(M_\star)\geq10.5$; {grey} open points).  Note that both sets of 
results decline similarly, implying that the evolution seen is not a symptom of Malmquist bias.  
{Vertical error bars are dominated by the assumed systematic uncertainty of 25\% in $L_\text{TIR}$ 
(described in the text), but also include the $1\sigma$ error on the 1.4-GHz flux.  Horizontal bars 
denote the full widths of the bins.
The star at $z=0$ represents the median $q_\text{TIR}=2.64\pm0.02$ from \citet{bell_estimating_2003}.
Our results appear to be systematically higher than this value, except where they decline at $z>1$, 
potentially as a result of our assumptions about the SEDs.}
}

\label{fig:q_tir}
\end{center}
\end{figure}

In Fig.~\ref{fig:q_tir} we plot the $q$ indices calculated from $L_\text{TIR}$
(listed in Table~\ref{tab:lum}) using Equation~\ref{eqn:q_fir} \citep{helou_thermal_1985}:  
\begin{equation}
q_\text{FIR} = \log{\left( \dfrac{L_\text{FIR}/3.75\times 10^{12}}{\text{W}} \right) } - \log{\left( \dfrac{L_\text{1.4~GHz}}{\text{W Hz}^{-1}}\right) }
\label{eqn:q_fir}
\end{equation}
Here we substitute $L_\text{TIR}$ for $L_\text{FIR} = L_{40-120\mum}$, (as in
\citealt{bell_estimating_2003} and \citealt{ivison_blast:far-infrared/radio_2009}, 
for example), and this difference should be noted when comparing to other work.
As an indication, the ratio of $L_\text{TIR}/L_\text{FIR}$ in the 
M51 template is $2.1$ (which implies $q_\text{TIR}-q_\text{FIR}=0.32$), 
but this ratio is likely to be variable since 
much of the longer wavelength emission can include contributions from dust
heated by older stellar populations \citep[as discussed for example
by][]{bell_estimating_2003}.

{The results for $q_\text{TIR}$ are shown in Fig.~\ref{fig:q_tir}
alongside the median result of \citet{bell_estimating_2003}
of $q_\text{TIR}=2.64\pm0.02$ for a FIR+FUV-selected sample of star-forming galaxies
at $z\approx0.0$.  We see that our results are generally a little 
higher than this value at $z<1$, and it is only due to an apparent evolution in our results
that they are more in agreement at high redshift.  The slight discrepancy is just within the 
errors allowed by our TIR normalization, and is likely to result from a systematic difference 
in the assumptions made about the SEDs and the associated calibration of TIR.  

The slight decline in our $q_\text{TIR}$ values
with redshift is described by} an error-weighted least-squares fit given by 
$q_\text{TIR}\propto(1+z)^\gamma$ where $\gamma=-0.11\pm0.07$.  Note that
stacking with the mass limit $\log(M_\star)\geq 10.5$ gives very
similar results, fit by the index $\gamma=-0.18\pm0.10$.
In comparison, the 24\mum\ sample of \citet{ivison_blast:far-infrared/radio_2009}
showed evidence for evolution over redshifts from 0 to 3, with an error-weighted 
least-squares fit of the same form given by $\gamma=-0.15\pm0.03$.  
Most recently, \citet{ivison_far-infrared/radio_2010} showed that a sample of
LIRGS detected by \textit{Spitzer} and stacked into \textit{Herschel} imaging 
at 100, 160, 250, 350 and 500\mum\ appear to exhibit an evolution in $q_\text{TIR}$
over $z=0-2$, with $\gamma=-0.04\pm0.03$ (or $-0.26\pm0.07$, discounting their 16 galaxies at
$z<0.5$ which were poorly matched in $L_\text{TIR}$ to the higher-redshift bins).

A slight decline of a similar scale ($\sim0.35$ dex) in $q_\text{TIR}$ with redshift ($0<z<1.4$) 
was also observed by \citet{sargent_vla-cosmos_2010}, in the median {IR/radio ratios} of their sample 
jointly selected in the IR and radio.  However this was at low ($2\sigma$) significance and 
the possibility of intrinsic evolution was rejected by the authors because the 
median at $z\sim1.4$ was within the scatter of their low-$z$ value, and moreover 
because the average at $z>2.5$ was very similar to the local value.
Instead they considered that their sample was more contaminated by AGN at increasing 
redshifts, and that the hot dust in these AGN caused $q_{24}$ ratios to remain constant,
while lower abundances of cold dust caused $q_{70}$ and $q_\text{TIR}$ to fall.  It is interesting 
to note that we similarly observe constant $q_{24}$ and falling $q_{70}$ and $q_\text{TIR}$,
but our observation of constant $q_{160}$ defies a similar explanation.  

In a second paper, \citet{sargent_no_evolution_2010} extended their earlier work using two 
volume-limited subsets of the joint sample: ULIRGs, and sources populating the bright end 
of the luminosity function defined by \citet{magnelli_0.4_2009}.
They showed that for both of these IR-bright populations, $q_\text{TIR}$
was constant out to redshift 1.4.  Following a correction for increased scatter
in their data beyond this redshift, they concluded that it remained constant
out to redshift 2. \footnote{We note that our data are not affected by the bias in $q$ described
by Sargent et al. (2010a,b) due to our selection in the IRAC bands.}  This result disagrees with 
that of \citet{ivison_blast:far-infrared/radio_2009}, which is flux-limited as opposed 
to volume-limited, showing the potential importance of selection effects.

The decline in our values of $q_\text{TIR}$ could still be caused by the
same effect that introduces the decline in $q_{70}$, since the template 
would underestimate both if the true SEDs were steeper at $\lambda\lesssim 
70\mum$.  Alternatively if $q_\text{TIR}$ really declines at high redshift then
something must be causing galaxies to emit less in the IR relative to the 
radio at increasing redshifts.  This could mean either a reduction in optical
depth, causing more UV photons to escape, or an increase in the confinement
and/or reprocessing efficiency of CR electrons leading to stronger radio emission.
This latter possibility cannot be ignored in the light of our observation that
radio spectral indices steepen at redshifts $z\gtrsim 1$, since \citet{lacki_physics_2009-1}
predict that steeper radio spectra are a sign of increasing electron calorimetry in normal galaxies.

In spite of these considerations, we remind the reader that the evolution in $q_\text{TIR}$ 
is at low significance (similar to that of \citealp{sargent_vla-cosmos_2010}), and our data are consistent
within $1.5\sigma$ with a non-evolution.  There is also the potential for some bias introduced by the variation
of spectral index with redshift: applying a constant spectral index of $-0.74$ to the radio $K$--corrections 
reduces {the $q_\text{TIR}$ evolution} to a level that is indistinguishable from being {constant: $\gamma=0.03\pm0.07$}).
{The evolution in $q_{70}$ however is not fully removed by this change.  Using the measured spectral indices
in each redshift bin we fit $q_\text{70}\propto(1+z)^{\gamma_{70}}$ with $\gamma_{70}=-0.15\pm0.05$, while using 
the constant spectral index we find $\gamma_{70}=-0.10\pm0.05$.  Nevertheless,} it is emphasised that the
measured spectral indices should give the most accurate $K$--correction, and Fig.~\ref{fig:alpha} shows that the
overall median of $-0.74$ is certainly not appropriate to represent the flux ratios in all of the bins.

\section{Conclusions}
We have studied the FRC as a function of redshift for NIR-selected
massive galaxies in the ECDFS, a sample which is unbiased by star-formation
activity.  We used a stacking analysis to evaluate the ratios of median 
FIR/radio fluxes of all galaxies in the sample, divided into redshift bins.  
This technique traces the typical objects in
the population of massive galaxies from low redshift back to their
formation epoch.  A thorough analysis of clustering
of the sample was used to correct for the differential effects of
confusion in the three FIR bands.  $K$--corrections were derived in the
radio and FIR using ratios of observed fluxes, ensuring as much as
possible a self-consistent analysis.  A mass-limited sub-sample was
also stacked to confirm the robustness of the results to Malmquist
bias.

The results for $q_{24}$, $q_{70}$ and $q_{160}$ show a slight decline
in the \textit{observed} relations, not dissimilar to the results of
previous studies, which can be largely accounted for by the FIR
$K$--correction using an M51 template.  After $K$--correction $q_{70}$ is
the only monochromatic index to show signs of evolution, suggesting that the 70-\mum\
$K$--correction may be less effective as a result of a steep slope in the
SED from $\sim 25-35\mum$ (corresponding to $z\sim1-2$) compared with M51.

Observed MIPS colours at all redshifts are more consistent with the
M51 template compared with hotter starburst galaxy templates,
indicating that the typical IR SEDs of stellar-mass-selected galaxies
at redshifts up to $\sim 0.8$ (at least) appear to be dominated by
cold dust.  At higher redshifts it is not possible to constrain the
dust temperature with MIPS colours, although it is still clear that
M51 is the closest template.  In contrast to this, both radio and
total IR luminosities rise significantly with increasing redshift, as
do derived SFRs.  Specific SFRs similarly rise steeply, in
agreement with {results in the literature 
\citep{cowie_new_1996,madau_star_1998,brinchmann_mass_2000,
bauer_specific_2005,feulner_connection_2005,prez-gonzlez_stellar_2008,
dunne_star_2009,damen_evolution_2009,pannella_star_2009,oliver_specific_2010}.}

The stacked radio data reveal tentative evidence for an evolution in
radio spectral index across the redshift range, an unexpected result that
implies some change in the radio loss processes in our sample towards
higher redshifts.  The most likely
explanation seems to be a shift towards greater inverse-Compton losses
of the CR electrons at $z>1$, {supporting}
the predictions of \citet{lacki_physics_2009}.

Overall our results show evidence that the FRC, measured from 24-\mum\ fluxes
or 160-\mum\ fluxes closer to the FIR peak, remains roughly
constant up to $z\sim2$, corresponding to 10 Gyr of cosmic time.  This
is similar to the conclusions of recent studies including
\citet{ibar_exploringinfrared/radio_2008, garn_relationship_2009,
younger_millimetre_2009, ivison_blast:far-infrared/radio_2009} and
\citet{sargent_vla-cosmos_2010}.  The issue is clouded however by measurements 
at 70\mum, which appear to show a declining $q$ index with redshift, and when
combined into a total IR luminosity, likewise show a slight decline (at low
significance).  This most likely implies a steeper spectral slope at wavelengths
around $25-35$\mum\ (compared with the M51 template), leading to insufficient
70-\mum\ $K$--corrections.  But a true evolution in the ratios of 70-\mum/radio luminosity
and of TIR/radio luminosity is plausible, considering the apparent
increase in electron-calorimetry behaviour at $z>1$, and considering the
fact that rest-frame 24, 70 and 160-\mum\ fluxes can arise from different components
of the dust in a galaxy.  It is also consistent with the results of 
\citet{seymour_investigatingfar-ir/radio_2009} for $q_{70}$ and 
\citeauthor{ivison_blast:far-infrared/radio_2009} (2010a/b) using BLAST/Herschel 
and \textit{Spitzer} observations to measure $q_\text{TIR}$.

Constraining the FIR SED is one of the greatest problems in understanding the FRC and the dust emission in general
from high redshift star-forming galaxies.  Upcoming surveys with \textit{Herschel}, such as
the \textit{Herschel} Multi-tiered Extragalactic Survey \citep[HerMES;][]{oliver_herschel_2010} 
and the \textit{Herschel} Astrophysical Terahertz Large Area Survey \citep[H-ATLAS;][]{eales_herschel_2010}
are anticipated to revolutionise our understanding of these topics by providing deep and
wide observations spanning the peak of FIR emission across the history of cosmic star formation.

\section*{Acknowledgements}
The authors are grateful to Jacqueline Monkiewicz of the National Optical Astronomy Observatory, Tucson, and Miwa Block of the University of Arizona, for reducing the 24-\mum\ and 160-\mum\ data.  NB acknowledges travel grants from the Institute of Physics and Royal Astronomical Society which made this work possible.  NB {also} wishes to thank Edo Ibar for his patient assistance with the calibration of $K$--corrections.  {Finally the authors would like to thank the anonymous referee for a thorough and insightful critique of the paper.}

\bibliographystyle{mn2e}
\bibliography{FIRRC2010.bib}

\clearpage
\begin{table*}
\begin{center}
\begin{tabular}{ c c c l c c c c c c c c c c }
\hline
Bin & $\langle z\rangle$ & $N_\text{stack}$ & Band & $S_\text{obs}$, $\mu$Jy & $\pm \sigma_{N}$, $\mu$Jy & $\pm \sigma_{S}$, $\mu$Jy & $S/N$ & $q_\text{corr} $ & $\pm \sigma_{S}$ & $S_{k}$, $\mu$Jy & $\pm \sigma_{S}$, $\mu$Jy & $q_{k}$ & $\pm \sigma_{S}$\\
(1) &  (2) &  (3) &     (4) &    (5) &    (6) &   (7) &  (8) &  (9) &   (10) &   (11) &  (12) & (13) & (14) \\ \hline
ALL & 0.73 & 3172 &    24\mum &  145.0 &    1.3 &    4.8 & 111.5 & 0.97 & 0.02 &  254.6 &    7.7 & 1.47 & 0.03  \\
 .  &      &      &    70\mum & 1037.2 &   21.9 &   34.3 &  47.3 & 1.85 & 0.02 & 2117.6 &   87.1 & 2.20 & 0.03  \\
 .  &      &      &   160\mum & 8058.3 &  378.1 &  357.4 &  21.3 & 2.63 & 0.03 & 5031.5 &  336.2 & 2.51 & 0.04  \\
 .  &      &      &  1.4~GHz  &   13.2 &    0.2 &    0.4 &  54.6 &      &      &   11.9 &    0.3 &      &       \\
 .  &      &      &  610~MHz  &   24.4 &    1.3 &    1.4 &  18.1 &      &      &   21.3 &    1.1 &      &       \\ \hline 
ZB0 & 0.21 &  528 &    24\mum &  186.4 &    3.1 &    9.5 &  59.6 & 1.09 & 0.05 &  222.3 &   14.2 & 1.44 & 0.07  \\
 .  &      &      &    70\mum & 1674.3 &   54.3 &   72.7 &  30.8 & 2.07 & 0.05 & 2068.3 &  159.9 & 2.34 & 0.08  \\
 .  &      &      &   160\mum & 9894.5 &  931.5 & 1023.8 &  10.6 & 2.73 & 0.07 & 5812.8 &  793.5 & 2.71 & 0.10  \\
 .  &      &      &  1.4~GHz  &   12.8 &    0.6 &    1.2 &  21.5 &      &      &   11.4 &    1.2 &      &       \\
 .  &      &      &  610~MHz  &   22.2 &    3.3 &    2.8 &   6.8 &      &      &   20.6 &    2.8 &      &       \\ \hline 
ZB1 & 0.53 &  528 &    24\mum &  142.3 &    3.1 &    9.1 &  45.5 & 0.94 & 0.05 &  235.1 &   15.1 & 1.31 & 0.04  \\
 .  &      &      &    70\mum & 1318.9 &   53.7 &  122.4 &  24.5 & 1.93 & 0.05 & 2305.5 &  242.4 & 2.29 & 0.04  \\
 .  &      &      &   160\mum & 9235.8 &  946.8 & 1056.6 &   9.8 & 2.67 & 0.07 & 4981.4 &  674.9 & 2.63 & 0.05  \\
 .  &      &      &  1.4~GHz  &   14.0 &    0.6 &    1.0 &  23.5 &      &      &   12.9 &    0.8 &      &       \\
 .  &      &      &  610~MHz  &   22.8 &    3.3 &    2.5 &   6.9 &      &      &   19.5 &    2.1 &      &       \\ \hline 
ZB2 & 0.67 &  529 &    24\mum &  144.3 &    3.1 &   10.2 &  46.1 & 0.97 & 0.05 &  251.6 &   20.1 & 1.35 & 0.04  \\
 .  &      &      &    70\mum & 1120.2 &   53.6 &  125.7 &  20.9 & 1.88 & 0.06 & 2212.5 &  253.3 & 2.29 & 0.05  \\
 .  &      &      &   160\mum & 7502.0 &  919.8 &  694.2 &   8.2 & 2.60 & 0.07 & 4152.7 &  629.9 & 2.59 & 0.08  \\
 .  &      &      &  1.4~GHz  &   13.2 &    0.6 &    0.7 &  22.2 &      &      &   10.9 &    0.6 &      &       \\
 .  &      &      &  610~MHz  &   20.5 &    3.3 &    2.8 &   6.2 &      &      &   16.4 &    2.2 &      &       \\ \hline 
ZB3 & 0.87 &  529 &    24\mum &  150.5 &    3.1 &   11.0 &  48.1 & 1.03 & 0.05 &  259.2 &   21.2 & 1.47 & 0.04  \\
 .  &      &      &    70\mum &  996.9 &   53.4 &  107.8 &  18.7 & 1.87 & 0.05 & 2262.5 &  246.8 & 2.37 & 0.05  \\
 .  &      &      &   160\mum & 6505.4 &  932.5 &  851.5 &   7.0 & 2.58 & 0.08 & 3801.9 &  723.8 & 2.63 & 0.08  \\
 .  &      &      &  1.4~GHz  &   12.1 &    0.6 &    0.6 &  20.3 &      &      &    9.2 &    0.6 &      &       \\
 .  &      &      &  610~MHz  &   18.9 &    3.3 &    4.0 &   5.8 &      &      &   14.2 &    2.9 &      &       \\ \hline 
ZB4 & 1.06 &  529 &    24\mum &  126.1 &    3.1 &    9.0 &  40.3 & 0.93 & 0.06 &  262.6 &   23.9 & 1.32 & 0.05  \\
 .  &      &      &    70\mum &  834.9 &   53.4 &   90.2 &  15.6 & 1.78 & 0.06 & 2108.4 &  218.9 & 2.23 & 0.06  \\
 .  &      &      &   160\mum & 7843.6 &  914.1 &  711.8 &   8.6 & 2.64 & 0.07 & 5288.1 &  625.9 & 2.60 & 0.07  \\
 .  &      &      &  1.4~GHz  &   12.5 &    0.6 &    1.1 &  21.1 &      &      &   12.9 &    1.1 &      &       \\
 .  &      &      &  610~MHz  &   25.7 &    3.3 &    2.8 &   7.8 &      &      &   25.1 &    2.6 &      &       \\ \hline 
ZB5 & 1.29 &  265 &    24\mum &  110.5 &    4.4 &    9.8 &  25.0 & 0.77 & 0.06 &  286.3 &   27.7 & 1.29 & 0.05  \\
 .  &      &      &    70\mum &  741.0 &   75.5 &  124.8 &   9.8 & 1.62 & 0.08 & 1976.6 &  354.0 & 2.12 & 0.08  \\
 .  &      &      &   160\mum & 8484.4 & 1287.5 & 1102.0 &   6.6 & 2.57 & 0.07 & 6713.7 & 1010.8 & 2.64 & 0.08  \\
 .  &      &      &  1.4~GHz  &   16.0 &    0.8 &    1.1 &  19.1 &      &      &   14.7 &    1.1 &      &       \\
 .  &      &      &  610~MHz  &   33.0 &    4.6 &    3.7 &   7.1 &      &      &   30.7 &    3.4 &      &       \\ \hline 
ZB6 & 1.61 &  264 &    24\mum &  168.2 &    4.4 &   17.7 &  38.0 & 0.95 & 0.06 &  325.3 &   24.7 & 1.20 & 0.05  \\
 .  &      &      &    70\mum &  539.0 &   75.9 &  101.2 &   7.1 & 1.48 & 0.09 & 1489.7 &  278.4 & 1.81 & 0.10  \\
 .  &      &      &   160\mum & 7404.0 & 1295.4 & 1205.1 &   5.7 & 2.50 & 0.09 & 7458.2 & 1078.3 & 2.56 & 0.10  \\
 .  &      &      &  1.4~GHz  &   16.2 &    0.8 &    1.2 &  19.3 &      &      &   20.3 &    1.8 &      &       \\
 .  &      &      &  610~MHz  &   40.8 &    4.7 &    3.5 &   8.8 &      &      &   48.5 &    4.2 &      &       \\ \hline 
\end{tabular}
\end{center}
\caption{Summary of stacking results.  Columns are as follows: (1) Redshift bin; (2) median redshift; 
(3) number of objects in stack; (4) Band; (5) Median observed flux before any corrections; 
(6) Measured $1\sigma$ noise (reduced by $\sqrt{N}$); 
(7) Statistical $1\sigma$ uncertainty on median flux \citep[following][]{gott_median_2001};
(8) Signal-to-noise ratio ($S_\text{obs}/\sigma_{N}$); 
(9) $q$ index for IR band after clustering correction; 
(10) Error on $q$ (using statistical uncertainty from column 7); 
(11) Median flux following clustering-- and $K$--corrections (using M51 template for MIPS and the measured $\alpha(z)$ for radio); 
(12) Statistical uncertainty on corrected flux (as in column 7); 
(13) $q$ index after $K$--corrections; 
(14) Statistical uncertainty on $K$--corrected $q$.}
\label{tab:stackresults}
\end{table*}

\end{document}